\documentclass{elsart}
\usepackage{natbib}
\usepackage{amsmath}
\usepackage{amssymb}
\usepackage{graphics}
\usepackage{amsbsy}
\journal{Physica A}

\usepackage{defs}
\usepackage{localdefs}

\begin{document}
\begin{frontmatter}

\title
{Many-particle hydrodynamic interactions in parallel-wall geometry: 
Cartesian-representation method}

\author{S. Bhattacharya},
\author{J. Blawzdziewicz},\and
\author{E. Wajnryb\thanksref{IPPT}}
\address{Department of Mechanical Engineering, Yale University, New
Haven, CT 06520-8286, USA}
\thanks[IPPT]{On leave from IPPT Warsaw, Poland}

\begin{abstract}

This paper describes the results of our theoretical and numerical
studies of hydrodynamic interactions in a suspension of spherical
particles confined between two parallel planar walls, under
creeping-flow conditions.  We propose a novel algorithm for accurate
evaluation of the many-particle friction matrix in this system---no
such algorithm has been available so far.

Our approach involves expanding the fluid velocity field into
spherical and Cartesian fundamental sets of Stokes flows.  The
interaction of the fluid with the particles is described using the
spherical basis fields; the flow scattered with the walls is expressed
in terms of the Cartesian fundamental solutions.  At the core of our
method are transformation relations between the spherical and
Cartesian basis sets.  These transformations allow us to describe the
flow field in a system that involves both the walls and particles.

We used our accurate numerical results to test the single-wall
superposition approximation for the hydrodynamic friction matrix.  The
approximation yields fair results for quantities dominated by single
particle contributions, but it fails to describe collective phenomena,
such as a large transverse resistance coefficient for linear arrays of
spheres.

\end{abstract}

\end{frontmatter}

\section{Introduction}
\label{Introduction}

Equilibrium and nonequilibrium behavior of colloidal suspensions in confined
geometries has recently been extensively discussed.  Examples of recent papers
include experimental studies of particle deposition on chemically patterned
planar walls
\cite{Lin-Crocker-Prasad-Schofield-Weitz-Lubensky-Yodh:2000},
investigations of collective dynamics in quasi-bidimensional
suspensions in slit pores
\cite{%
Acuna_Campa-Carbajal_Tinoco-Arauz_Lara-Medina_Noyola:1998,%
Pesche-Kollmann-Nagele:2001,%
Santana_Solano-Arauz_Lara:2002%
}, 
and observations of drainage behavior of particle-stabilized thin
liquid films
\cite{Sethumadhavan-Nikolov-Wasan:2001}.
The research has been stimulated, in part, by emerging applications---such as
microfluidic devices and production of photonic materials by self-assembly of
colloidal crystals
\cite{ Subramanian-Manoharan-Thorne-Pine:1999,%
Seelig-Tang-Yamilov-Cao-Chang:2002%
}.
The investigations have also been considerably influenced by development of
new experimental techniques, including the evanescent-wave microscopy
\cite{%
Prieve-Luo-Lanni:1987,%
Walz-Suresh:1995%
},
computerized video microscopy
\cite{%
Faucheux-Libchaber:1994,%
Lin-Yu-Rice:2000,%
Santana_Solano-Arauz_Lara:2002,%
Palberg-Biehl:2003%
}, 
and optical tweezers \cite{Crocker-Matteo-Dinsmore-Yodh:1999}.

While the equilibrium structure of confined colloidal suspensions is
fully determined by the particle--wall and interparticle interaction
potentials, the dynamics is also significantly affected by the
many-body hydrodynamic forces.  The effect of the hydrodynamic
interactions on particle motion can be expressed in terms of the
$N$-particle friction and mobility matrices
\cite{Kim-Karrila:1991}, which depend on the particle positions
and the wall geometry.

For spherical particles in an unbounded space, efficient algorithms
for evaluation of the friction and mobility matrices have been
developed
\cite{
Durlofsky-Brady-Bossis:1987,%
Ladd:1988,%
Cichocki-Felderhof-Hinsen-Wajnryb-Blawzdziewicz:1994,%
Sangani-Mo:1996,%
Sierou-Brady:2001%
}.
The algorithms combine multipolar expansion methods with the
lubrication approximation for particles in close proximity
\cite{Durlofsky-Brady-Bossis:1987}.  This approach has recently
been generalized by Cichocki
\etal~\cite{Cichocki-Jones:1998,Cichocki-Jones-Kutteh-Wajnryb:2000} to
systems of particles bounded by a single planar wall; the
particle--wall hydrodynamic interactions were included using the image
representation of the flow reflected from the wall
\cite{Cichocki-Jones:1998}.

Much less progress has been made for suspensions confined between two planar
walls (e.g, in a slit pore, or between two glass plates).  For a single
particle, several {\it ad hoc} approximations for the mobility matrix have
been proposed \cite{Lobry-Ostrowsky:1996,Benesch-Yiacoumi-Tsouris:2003}, and
numerical results obtained by boundary-integral methods are available
\cite{%
Ganatos-Weinbaum-Pfeffer:1980,%
Ganatos-Pfeffer-Weinbaum:1980,%
Staben-Zinchenko-Davis:2003%
}.
Recently, we have developed an exact image representation of the flow
between two walls \cite{Bhattacharya-Blawzdziewicz:2002}, which allows
accurate evaluation of the single-particle friction matrix by a
multipolar expansion technique.  However, none of the above methods
has been generalized to multiparticle systems, due to a large
numerical cost of boundary-integral calculations or the slow
convergence of the image solutions.

Two extensions of the free-space Stokesian-dynamics algorithm
\cite{Durlofsky-Brady-Bossis:1987} to wall bounded systems have been
proposed by Brady and his collaborators
\cite{Durlofsky-Brady:1989,Nott-Brady:1994,Morris-Brady:1998}. In the
first approach the walls are discretized \cite{Durlofsky-Brady:1989},
and in the second they are modeled as static, closely packed arrays of
spheres \cite{Nott-Brady:1994,Morris-Brady:1998}.  The first method
has not been further explored.  The results obtained using the second
method are only qualitative, because the walls are porous and rough.

To overcome the above-mentioned problems, we adopt here an alternative
approach, based on Fourier analysis of the flow field reflected from
the walls.\footnote{Recently, investigations along similar lines have
also been reported by Jones \cite{Jones:2004,Jones:2004a}.}  According
to our method, the flow field in the system is expanded using two
basis sets of solutions of Stokes equations---the spherical and the
Cartesian basis.  The spherical basis is used for a description of the
flow field scattered from the particles, and the Cartesian basis is
appropriate for the wall geometry. The key result of our study is a
set of transformation formulas for conversion between the spherical
and Cartesian representations.  The transformation formulas allow to
evaluate the spherical matrix elements of the Green function for
Stokes flow in the presence of the walls in terms of simple
two-dimensional Fourier integrals.

The results of our theoretical analysis have been implemented in a
numerical procedure for evaluating multi-particle hydrodynamic
interactions in a suspension of spheres confined between two planar
walls.  The procedure combines the expansions of the flow field into
the spherical and Cartesian basis fields with the two-particle
superposition approximation for the friction matrix, in order to
include slowly convergent lubrication corrections.  Since the force
multipoles induced on particle surfaces are included to arbitrary
order, highly accurate results are obtained.

Examples of numerical results for two-particle and many-particle
systems are provided.  In particular, our results illustrate the role
of the far-field flow produced in the space between the walls by the
moving particles.  We show that the single-wall superposition
approximation does not correctly describe the far-field flow, and thus
it fails to capture some important collective phenomena such as the
increased hydrodynamic resistance due to the backflow produced by the
moving particles.

This paper is organized as follows.  In Section \ref{Induced-force
formulation} we summarize the induced-force formulation of the Stokes-flow
equations for a multiparticle system in the wall presence. In Section
\ref{Matrix representation} the induced-force equations are transformed into
an infinite array of algebraic equations for the induced-force multipoles,
using a multipolar expansion of Stokes flow.

Our main theoretical results are presented in Sections \ref{Cartesian
basis}--\ref{Suspension between two planar walls--matrix elements}.
In Section \ref{Cartesian basis} the Cartesian basis sets of Stokes flows
are defined, and the transformation formulas for conversion between
the Cartesian and the spherical multipolar bases sets are derived.
The displacement and conversion formulas are then used to obtain
two-dimensional Fourier representations of the matrix elements of
Green operator for infinite space (Section \ref{Fourier representation of
displacement matrix in spherical basis}), halfspace bounded by a
single wall (Section \ref{Hyrodynamic interactions with single planar
wall}), and a region bounded by two parallel planar walls (Section
\ref{Suspension between two planar walls--matrix elements}).

A numerical algorithm for computation of hydrodynamic interactions in
a suspension of spheres confined in a region bounded by two parallel
walls is described in Section \ref{Numerical implementation}.
Numerical examples of the friction matrix, evaluated using this
algorithm, are given in Section \ref{Results}.  Directions for further
development of our method are indicated in the concluding Section
\ref{Conclusions}.  Some technical details are presented in Appendices
\ref{Spherical basis}--\ref{Large k behavior of integrands delta Psi}.

\section{Induced-force formulation}
\label{Induced-force formulation}

We consider a suspension of $N$ spherical particles of radius $a$
moving in an incompressible Newtonian fluid of viscosity $\eta$.  The
suspension is bounded by a single planar wall or two parallel planar
walls.  The creeping-flow conditions are assumed; therefore, the fluid
flow in the system depends only on the instantaneous particle
configuration and velocities.  The configuration is described by the
positions $(\bR_1,\ldots,\bR_N)$ of particle centers.  The
translational and rotational velocities of the particles are $\bU_i$
and $\bOmega_i$, where $i=1,\ldots,N.$

The effect of the suspended particles on the surrounding fluid can be
described in terms of the induced force distributions on the particle
surfaces
\begin{equation}
\label{induced forces}
\bF_i(\br)=a^{-2}\delta(r_i-a)\bff_i(\br),
\end{equation}
where 
\begin{equation}
\label{definition of r_i}
\br_i=\br-\bR_i
\end{equation}
and $r_i=|\br_i|$.  By definition of the induced force, the flow field
\begin{equation}
\label{flow field produced by induced forces}
\bv(\br)=\externalVelocity+\sum_{i=1}^N
  \int\bT(\br,\br')\bcdot\bF_i(\br')\diff\br'
\end{equation}
is identical to the velocity field in the presence of the particles
\cite{Cox-Brenner:1967,Mazur-Bedeaux:1974,Felderhof:1976b}.  In
the above equation, $\externalVelocity$ denotes the imposed flow, and
the integral term describes the flow generated by the induced forces.
Here
\begin{equation}
\label{Green's function}
\bT(\br,\br')=\bT_0(\br-\br')+\bT'(\br,\br')
\end{equation}
is the Green function for the Stokes flow in the presence of the
boundaries,
\begin{equation}
\label{Oseen tensor}
\bT_0(\br)=\frac{1}{8\upi\eta r}(\identityTensor+\hat\br\hat\br)
\end{equation}
denotes the Oseen tensor (where $\identityTensor$ is the identity
tensor, and $\hat\br=\br/r$), and $\bT'(\br,\br')$ describes the flow
reflected from the walls.

The induced force distribution $\bF_i$ on the surface of particle $i$
and the flow $\incidentVelocity{i}$ incident to this particle are
linearly related.  The relation can be expressed in the form
\begin{equation}
\label{definition of operator Z}
\bF_i=-\bZ_i(\incidentVelocity{i}-\bv_i^{\rm rb}),
%
%
\end{equation}
where
\begin{equation}
\label{rigid-body velocity of drop i}
\bv_i^{\rm rb}(\br)=\bU_i+\bOmega_i\times\br_i
\end{equation}
denotes the rigid-body velocity field corresponding to the particle
motion, and
\begin{equation}
\label{incoming flow in reference frame of particle}
\barincidentVelocity{i}=\incidentVelocity{i}-\bv_i^{\rm rb}
\end{equation}
is the incident flow in the reference frame moving with the particle.
The Stokes flow field \refeq{incoming flow in reference frame of
particle} is fully determined by its boundary value on the particle
surface $S_i$ and the condition that $\bar\bv_i^{\rm in}$ is
nonsingular in the region occupied by the particle.  Thus,
\refeq{definition of operator Z} can be interpreted as a linear
functional relation between the vector fields \refeq{induced forces}
and \refeq{incoming flow in reference frame of particle} specified on
the surface $S_i$.  Since a nonzero flow \refeq{incoming flow in
reference frame of particle} always produces a nonzero force
distribution $\bF_i$, relation \refeq{definition of operator Z} can be
inverted\begin{equation}
\label{inverse Z relation}
\incidentVelocity{i}-\bv_i^{\rm rb}
=-[\bZ_i^{-1}\bF_i](\br),\qquad\br\in S_i.
\end{equation}
For specific particle models, explicit expressions for the operator
$\bZ_i$ are obtained by solving Stokes equations for an isolated
particle subject to an external flow in an unbounded fluid
\cite{%
Jones-Schmitz:1988,%
Cichocki-Felderhof-Schmitz:1988,%
Blawzdziewicz-Wajnryb-Loewenberg:1999%
}.  

The flow $\incidentVelocity{i}$ incident to a particle $i$ in a
multiparticle system is defined by the equation
\begin{equation}
\label{incident flow}
\bv(\br)=
\incidentVelocity{i}(\br)+\scatteredVelocity{i}(\br),
\end{equation}
where $\bv(\br)$ is the total flow \refeq{flow field produced by
induced forces}, and 
\begin{equation}
\label{flow scattered by particle}
\scatteredVelocity{i}(\br)
   =\int\bT_0(\br-\br')\bcdot\bF_i(\br')\diff\br'
\end{equation}
represents the flow scattered by the considered particle.  By
collecting relations \refeq{inverse Z relation}--\refeq{flow scattered
by particle} we obtain the expression
\begin{equation}
\label{expression relating Z inverse and self Oseen term}
\bv(\br)=\bv_i^{\rm rb}(\br)-[\bZ_i^{-1}\bF_i](\br)
   +\int\bT_0(\br-\br')\bcdot\bF_i(\br')\diff\br',
\qquad
   \br\in S_i,
\end{equation}
for the flow at the surface $S_i$ of the particle $i$.  For rigid
spheres, the velocity field $\bv(\br)$ in equation \refeq{expression
relating Z inverse and self Oseen term} satisfies the no-slip boundary
condition
\begin{equation}
\label{velocity boundary condition for rigid particle}
\bv(\br)=\bv_i^{\rm rb}(\br),\qquad\br\in S_i.
\end{equation}
Accordingly, we have the identity
\cite{Cichocki-Jones-Kutteh-Wajnryb:2000}
\begin{equation}
\label{inverse Z for rigid spheres}
[\bZ_i^{-1}\bF_i](\br)
   =\int\bT_0(\br-\br')\bcdot\bF_i(\br')\diff\br',
\qquad
   \br\in S_i
\end{equation}
for such particles.

By combining expressions \refeq{flow field produced by induced forces}
and \refeq{expression relating Z inverse and self Oseen term}, we
get the boundary-integral equation for the induced force densities
$\bF_i$,
\begin{eqnarray}
\label{boundary-integral equation for induced-force density}
[\bZ_i^{-1}\bF_i](\br)
   +\sum_{j=1}^N\int
      [(1-\delta_{ij})\bT_0(\br-\br')+\bT'(\br,\br')]
      \bcdot\bF_j(\br')\diff\br'
   =\bv_i^{\rm rb}(\br)-\externalVelocity(\br),&&
\nonumber\\\br\in S_i.\qquad&&
\end{eqnarray}
In the following sections, equation \refeq{boundary-integral equation
for induced-force density} is transformed into an infinite set of
algebraic equations for the multipole moments of the induced force,
with coefficients expressed in terms of two-dimensional Fourier
integrals.

\section{Matrix representation}
\label{Matrix representation}
\subsection{Spherical basis}
\label{spherical basis}

The matrix representation of equation \refeq{boundary-integral
equation for induced-force density} is obtained by expanding fluid
velocity fields into sets of fundamental solutions of Stokes equations
in spherical coordinates, and expressing the induced-force
distributions in terms of the corresponding force multipoles.  In our
analysis we employ sets of basis fields that are closely related to
the sets introduced by Cichocki
\etal~\cite{Cichocki-Felderhof-Schmitz:1988}; we use, however, a
different normalization to emphasize important symmetries of the
problem.

The singular and nonsingular basis sets of solutions of Stokes
equations $\sphericalBasisM{lm\sigma}(\br)$ and
$\sphericalBasisP{lm\sigma}(\br)$ (where $l=1,2,\ldots$;
$m=-l,\ldots,l$; and $\sigma=0,1,2$) are defined by the following
conditions: ({\it i}\/) the basis velocity fields are homogeneous
functions of the radial variable $r$,
\begin{equation}
\label{spherical basis v -}
\sphericalBasisM{lm\sigma}(\br)
       =\sphericalBasisCoefM{lm\sigma}(\theta,\phi)r^{-(l+\sigma)},
\end{equation}
\begin{equation}
\label{spherical basis v +}
\sphericalBasisP{lm\sigma}(\br)
       =\sphericalBasisCoefP{lm\sigma}(\theta,\phi)r^{l+\sigma-1},
\end{equation}
where $(r,\theta,\phi)$ represent the vector $\br$ in spherical
coordinates; ({\it ii}\/) the coefficients
$\sphericalBasisCoefM{lm\sigma}(\theta,\phi)$ and
$\sphericalBasisCoefP{lm\sigma}(\theta,\phi)$ are combinations of
vector spherical harmonics with angular order $l$ and azimuthal order
$m$; and ({\it iii}\/) the basis velocity fields
$\sphericalBasisPM{lm\sigma}(\br)$ satisfy the following hierarchies
of curl relations
\begin{subequations}
\label{spherical curl -}
\begin{equation}
\label{spherical curl - 1}
\sphericalBasisM{lm1}
   =-\im\boldsymbol{\nabla}\boldsymbol{\times}\sphericalBasisM{lm0},
\end{equation}
\begin{equation}
\label{curl - 2}
\sphericalBasisM{lm2}
   =-\im\boldsymbol{\nabla}\boldsymbol{\times}\sphericalBasisM{lm1},
\end{equation}
\end{subequations}
and
\begin{subequations}
\label{spherical curl +}
\begin{equation}
\label{curl + 1}
\sphericalBasisP{lm1}
   =\im\boldsymbol{\nabla}\boldsymbol{\times}\sphericalBasisP{lm2},
\end{equation}
\begin{equation}
\label{curl + 0}
\sphericalBasisP{lm0}
   =\im\boldsymbol{\nabla}\boldsymbol{\times}\sphericalBasisP{lm1}.
\end{equation}
\end{subequations}
The above identities imply that only the solutions
$\sphericalBasisM{lm0}$ and $\sphericalBasisP{lm2}$ have nonzero
corresponding pressure fields, and that the solutions
$\sphericalBasisM{lm2}$ and $\sphericalBasisP{lm0}$ represent
potential flows, i.e.,
\refstepcounter{equation}
$$
\label{spherical zero curl}
\bnabla\btimes\sphericalBasisM{lm2}
   =0,\qquad\bnabla\btimes\sphericalBasisP{lm0}=0.
\eqno{(\theequation{\mathit{a},\mathit{b}})}
$$ The proportionality coefficient in the curl relations \refeq{spherical curl
-} and \refeq{spherical curl +} is determined by the requirement that the
expansion of the Oseen tensor in basis functions \refeq{spherical
basis v -} and \refeq{spherical basis v +} has the form
\cite{Cichocki-Felderhof-Schmitz:1988,Perkins-Jones:1991}
\refstepcounter{equation}
$$
\label{expansion of Oseen tensor in spherical basis}
\eta\bT_0(\br-\br')=
\left\{
   \begin{array}{ll}
     \displaystyle\sum_{lm\sigma}
        \sphericalBasisM{lm\sigma}(\br)\sphericalBasisPcon{lm\sigma}(\br'),
        \qquad &r>r',\\&\\
     \displaystyle\sum_{lm\sigma}
        \sphericalBasisP{lm\sigma}(\br)\sphericalBasisMcon{lm\sigma}(\br'),
        \qquad &r<r'.
   \end{array}
\right.
\eqno{(\theequation{\mathit{a},\mathit{b}})}
$$ 
The conditions \refeq{spherical basis v -}--\refeq{expansion of Oseen
tensor in spherical basis} determine the basis fields
$\sphericalBasisPM{lm\sigma}$ up to a single normalization constant,
which is set by an additional requirement that
\begin{equation}
\label{additional normalization requirement}
\sphericalBasisP{lm0}=\boldsymbol{\nabla}r^lY_{lm},
\end{equation}
where $Y_{lm}$ is the normalized scalar spherical harmonic (as defined
by Edmonds \cite{Edmonds:1960}).

The flow fields \refeq{spherical basis v -} and \refeq{spherical basis
v +} form complete sets of singular and non-singular solutions of
Stokes equations in the representation appropriate for spherical
symmetry.  However, they do not form orthonormal sets with respect to
the natural functional scalar product for vector fields $\bA$ and
$\bB$ on the spherical surface $r=b$.  Following the approach of
Cichocki \etal~\cite{Cichocki-Felderhof-Schmitz:1988} we thus
introduce the reciprocal basis fields
$\reciprocalSphericalBasisPM{lm\sigma}$, which are defined by the
orthogonality relations
\begin{equation}
\label{orthogonality relations for spherical reciprocal basis}
\langle\deltab{b}\reciprocalSphericalBasisPM{lm\sigma}\mid
    \sphericalBasisPM{l'm'\sigma'}\rangle
    =\delta_{ll'}\delta_{mm'}\delta_{\sigma\sigma'}
\end{equation}
for all values of parameter $b>0$, where
\begin{equation}
\label{delta b}
\deltab{b}(\br)=b^{-2}\delta(r-b),
\end{equation}
and 
\begin{equation}
\label{scalar product}
\langle\bA\mid\bB\rangle=\int \bA^*(\br)\boldsymbol{\cdot}\bB(\br)\diff\br.
\end{equation}
 The functions
$\reciprocalSphericalBasisPM{lm\sigma}$ have a similar structure to
the functions $\sphericalBasisPM{lm\sigma}$, i.e., they have a
separable form
\begin{equation}
\label{spherical basis w -}
\reciprocalSphericalBasisM{lm\sigma}(\br)
       =\reciprocalSphericalBasisCoefM{lm\sigma}(\theta,\phi)r^{l+\sigma},
\end{equation}
\begin{equation}
\label{spherical basis w +}
\reciprocalSphericalBasisP{lm\sigma}(\br)
       =\reciprocalSphericalBasisCoefP{lm\sigma}(\theta,\phi)r^{-(l+\sigma-1)},
\end{equation}
with the coefficients
$\reciprocalSphericalBasisCoefPM{lm\sigma}(\theta,\phi)$ given by
combinations of vector spherical harmonics with angular order $l$ and
azimuthal order $m$.  Explicit relations for the functions
$\sphericalBasisCoefPM{lm\sigma}(\theta,\phi)$ and
$\reciprocalSphericalBasisCoefPM{lm\sigma}(\theta,\phi)$ in equations
\refeq{spherical basis v -}, \refeq{spherical basis v +},
\refeq{spherical basis w -}, and \refeq{spherical basis w +} are
listed in Appendix \ref{Spherical basis}.

\subsection{Equations for induced-force multipole moments}
\label{Induced-force multipole equations}

In the multipolar-representation method, the boundary-integral
equation \refeq{boundary-integral equation for induced-force density}
is transformed into an infinite set of linear algebraic equations for
the multipolar moments of the induced-force distributions
\refeq{induced forces}.  The multipolar expansion of the 
distribution $\bF_i$ is defined by the relation
\begin{equation}
\label{induced force in terms of multipoles}
\bF_i(\br)
   =\sum_{lm\sigma}
      f_i(lm\sigma)
            \deltab{a}(\br_i)
         \reciprocalSphericalBasisP{lm\sigma}(\br_i).
\end{equation}
The corresponding multipolar moments are given by 
\begin{eqnarray}
\label{induced force multipoles}
f_i(lm\sigma)
 &=&\int\sphericalBasisPcon{lm\sigma}(\br_i)
     \bcdot\bF_i(\br)\diff\br
\nonumber\\
 &=&\langle\sphericalBasisP{lm\sigma}(i)\mid\bF_i\rangle,
\end{eqnarray}
consistent with the orthogonality relation \refeq{orthogonality
relations for spherical reciprocal basis}.  In the above equation we
introduce the standard bra--ket notation, with an additional
convention that $|\bA\rangle$ represents the vector field $\bA(\br)$
and $|\bA(i)\rangle$ denotes $\bA(\br_i)$.

The linear algebraic equations for the multipolar moments of the
induced force \refeq{induced force multipoles} are obtained by
projecting the linear operators in the boundary-integral equation
\refeq{boundary-integral equation for induced-force density} onto the
reciprocal basis \refeq{spherical basis w +}.  The resulting matrix
representation of equation \refeq{boundary-integral equation for
induced-force density} can be written in the form
\begin{equation}
\label{induced force equations for matrix elements}
\sum_{j=1}^N\sum_{l'm'\sigma'}
   \GrandMobilityElement_{ij}(lm\sigma\mid l'm'\sigma')
      f_j(l'm'\sigma')
      =c_i(lm\sigma),
\end{equation}
where
\begin{equation}
\label{Grand Mobility matrix elements}
\GrandMobilityElement_{ij}(lm\sigma\mid l'm'\sigma')=
   Z_{ij}^{-1}(lm\sigma\mid l'm'\sigma')
   +
      \GreenFreeElement_{ij}(lm\sigma\mid l'm'\sigma')
   +
      \GreenWallElement_{ij}(lm\sigma\mid l'm'\sigma'),
\end{equation}
and
\begin{eqnarray}
\label{w projections of Z^-1}
&&Z_{ij}^{-1}(lm\sigma\mid l'm'\sigma')=
   \delta_{ij}
   \langle\deltab{a}(i)\reciprocalSphericalBasisP{lm\sigma}(i)
\mid
   \bZ_j^{-1}
\mid
    \deltab{a}(j)\reciprocalSphericalBasisP{l'm'\sigma'}(j)\rangle,
\\\nonumber\\
\label{w projections of Oseen tensor}
&&\GreenFreeElement_{ij}(lm\sigma\mid l'm'\sigma')
   =(1-\delta_{ij})
   \langle\deltab{a}(i)\reciprocalSphericalBasisP{lm\sigma}(i)
\mid
   \bT_0
\mid
   \deltab{a}(j)\reciprocalSphericalBasisP{l'm'\sigma'}(j)\rangle,
\\\nonumber\\
\label{w projections of perturbation Greens tensor}
&&\GreenWallElement_{ij}(lm\sigma\mid l'm'\sigma')
   =\langle\deltab{a}(i)\reciprocalSphericalBasisP{lm\sigma}(i)
\mid
   \bT'
\mid
   \deltab{a}(j)\reciprocalSphericalBasisP{l'm'\sigma'}(j)\rangle.
\end{eqnarray}
In equations \refeq{w projections of Z^-1}--\refeq{w projections of
perturbation Greens tensor} 
\begin{equation}
\label{integral operator}
[\bT\bb](\br)=\int\bT(\br,\br')\bcdot\bb(\br')\diff\br',
\end{equation}
and the bra--ket notation introduced in equation \refeq{induced force
multipoles} is used.  The matrix elements \refeq{w projections of
Oseen tensor} and \refeq{w projections of perturbation Greens tensor}
are independent of the particle radius $a$, because the orthogonality
relation \refeq{orthogonality relations for spherical reciprocal
basis} holds for all values of the parameter $b$.  The coefficients
$c_i(lm\sigma)$ on the right side of equation \refeq{induced force
equations for matrix elements} are defined by the expansion
\begin{equation}
\label{expansion of external flow}
\bv_i^{\rm rb}(\br)-\externalVelocity(\br)
   =\sum_{lm\sigma}c_i(lm\sigma)\sphericalBasisP{lm\sigma}(\br_i)
\end{equation}
of the imposed flow field relative to the rigid-body particle motion
\refeq{rigid-body velocity of drop i} into the basis functions
\refeq{spherical basis v +} centered at the position of particle $i$.
Inserting the above expression into the orthogonality relation
\refeq{orthogonality relations for spherical reciprocal basis} yields
\begin{equation}
\label{explicit formula for c}
   c_i(lm\sigma)=
\langle
   \deltab{a}(i)\reciprocalSphericalBasisP{lm\sigma}(i)
\mid
   \bv_i^{\rm rb}-\externalVelocity
\rangle.
\end{equation}

For a system of identical particles, the matrix elements \refeq{w
projections of Z^-1} of the one-particle operator $\bZ_i^{-1}$ are
independent of the particle label $i$.  Since the particles are
assumed to be spherical, the matrix elements \refeq{w projections of
Z^-1} are diagonal in the multipolar orders $l$ and $m$, and
independent of $m$,
\begin{equation}
\label{Z diagonal}
Z_{ij}^{-1}(lm\sigma\mid l'm'\sigma')
   =\delta_{ij}\delta_{ll'}\delta_{mm'}Z_i^{-1}(l;\sigma\mid\sigma').
\end{equation}
By specifying equation \refeq{induced force equations for matrix
elements} for a single isolated particle $i$ in an unbounded fluid and
using the diagonality relation \refeq{Z diagonal} we obtain the linear
formula
\begin{equation}
\label{one-particle induced force equation}
\sum_{\sigma'}Z_i^{-1}(l;\sigma\mid\sigma')f_i(lm\sigma')
   =c_i(lm\sigma).
\end{equation}
Inserting the Oseen tensor in the form (\ref{expansion of Oseen tensor
in spherical basis}\textit{a}) into equation \refeq{flow scattered by
particle} and using the definition \refeq{induced force multipoles} we
also get
\begin{equation}
\label{multipolar expansion of scattered field}
\scatteredVelocity{i}(\br)
   =\eta^{-1}\sum_{lm\sigma}
      f_i(lm\sigma)\sphericalBasisM{lm\sigma}(\br_i).
\end{equation}
According to the above relation, the multipolar moments
$f_i(lm\sigma)$ can be interpreted as the expansion coefficient of the
flow field $\scatteredVelocity{i}$ scattered by the particle $i$ into
the basis velocity fields \refeq{spherical basis v -}.  It follows
that the matrix $Z_i^{-1}(l;\sigma\mid\sigma')$ relates the expansion
coefficients of the incident and the scattered flows.  For hard
spheres, porous particles, spherical viscous drops, and spherical
drops covered by an incompressible surfactant layer, explicit
expressions for the matrix elements \refeq{Z diagonal} are known
\cite{%
Jones-Schmitz:1988,%
Cichocki-Felderhof-Schmitz:1988,%
Blawzdziewicz-Wajnryb-Loewenberg:1999%
}.  
(Note, however, a different normalization of the basis functions here
and in the above references, as discussed in Appendix \ref{Spherical
basis})

The matrix elements of the free-space Oseen operator \refeq{w
projections of Oseen tensor} are also known, since they are simply
linked to the elements of the displacement matrix $S^{+-}$ that was
evaluated by Felderhof and Jones~\cite{Felderhof-Jones:1989}.  To show
this relation we insert expression (\ref{expansion of Oseen tensor in
spherical basis}\textit{a}), specified for $\bT_0(\br_j-\br'_j)$ with
$\br_j=\br-\bR_j$ and $\br'_j=\br'-\bR_j$, into equation \refeq{w
projections of Oseen tensor}, and use the orthogonality condition
\refeq{orthogonality relations for spherical reciprocal basis} for the
fields centered at the position of the particle $j$.  As the result we
find \cite{Cichocki-Jones-Kutteh-Wajnryb:2000}
\begin{equation}
\label{spherical displacement expression for Oseen matrix elements}
\GreenFreeElement_{ij}(lm\sigma\mid l'm'\sigma')
   =\eta^{-1}\langle\deltab{a}(i)\reciprocalSphericalBasisP{lm\sigma}(i)
\mid
   \sphericalBasisM{l'm'\sigma'}(j)\rangle.
\end{equation}
The matrix element on the right side of the above equation corresponds
to the expansion of the singular flow field
$\sphericalBasisM{l'm'\sigma'}$ centered at the position of the
particle $j$ into the nonsingular basis flow fields
$\sphericalBasisP{lm\sigma}$ centered at the position of the particle
$i$,
\begin{equation}
\label{displacement of spherical flow fields}
\sphericalBasisM{l'm'\sigma'}(\br_j)
   =\sum_{lm\sigma}\sphericalBasisP{lm\sigma}(\br_i)
\langle\deltab{a}(i)\reciprocalSphericalBasisP{lm\sigma}(i)
\mid
   \sphericalBasisM{l'm'\sigma'}(j)\rangle.
\end{equation}
According to the definition of the displacement matrix
\cite{Felderhof-Jones:1989} we thus have
\begin{equation}
\label{relation of Oseen matrix  to spherical displacement}
\GreenFreeElement_{ij}(lm\sigma\mid l'm'\sigma')
   =\eta^{-1}
   \sphericalDisplacementElements{+-}(\bR_i-\bR_j;lm\sigma\mid l'm'\sigma').
\end{equation}
We note that the displacement matrix
$\sphericalDisplacementElements{+-}$, introduced above, is normalized
differently than the matrix $S^{+-}$ defined by Felderhof and Jones
\cite{Felderhof-Jones:1989} (cf., the transformation \refeq{relation
between BBW and CFS bases} between the corresponding basis fields.)

As a result of the Lorentz symmetry
\begin{equation}
\label{symetry of Green functions}
\bT_\alpha(\br,\br')=\bT^\dagger_\alpha(\br',\br)
\end{equation}
of the Green functions $\bT_\alpha=\bT_0,\bT'$ (where the dagger
denotes the transpose of the tensor) and the symmetry of the scalar
product \refeq{scalar product}, the matrix elements \refeq{w
projections of Oseen tensor} and \refeq{w projections of perturbation
Greens tensor} satisfy the reciprocal relations
\begin{eqnarray}
\label{reciprocal relations elements of Oseen operator}
\GreenFreeElement_{ij}(lm\sigma\mid l'm'\sigma')
   &=&\GreenFreeElementCon_{ji}(l'm'\sigma'\mid lm\sigma),
\\
\label{reciprocal relations elements wall Green operator}
\GreenWallElement_{ij}(lm\sigma\mid l'm'\sigma')
   &=&\GreenWallElementCon_{ji}(l'm'\sigma'\mid lm\sigma),
\end{eqnarray}
where the asterisk denotes the complex conjugate.  The matrix elements
\refeq{Z diagonal} of the one-particle scattering operator have a
similar symmetry,
\begin{equation}
\label{reciprocal relations elements of one-particle scattering operator}
Z_i^{-1}(l;\sigma\mid\sigma')=Z_i^{-1}(l;\sigma'\mid\sigma).
\end{equation}
The matrix elements \refeq{reciprocal relations elements of
one-particle scattering operator} are real, due to the diagonality
\refeq{Z diagonal} of the matrix \refeq{w projections of Z^-1} in the
azimuthal number $m$.

\subsection{Matrix notation}
\label{Matrix notation}

In what follows we will use a compact matrix notation in the
three-dimensional linear space with the components corresponding to
the indices $\sigma=0,1,2$ that identify the tensorial character of
the basis flow fields \refeq{spherical basis v -} and \refeq{spherical
basis v +}.  Accordingly, the matrices with the elements \refeq{w
projections of Z^-1}--\refeq{w projections of perturbation Greens
tensor} will be denoted by \mbox{$\Zsingle_{ij}^{-1}(lm\mid l'm')$},
$\GreenFree_{ij}(lm\mid l'm')$, and $\GreenWall_{ij}(lm\mid l'm')$,
respectively; the matrices with the elements
$Z_i^{-1}(l;\sigma\mid\sigma')$ and
$\sphericalDisplacementElements{+-}(\bR_i-\bR_j;lm\sigma\mid
l'm'\sigma')$ will be denoted by $\Zsingle_i^{-1}(l)$ and
$\sphericalDisplacement{+-}(\bR_i-\bR_j;lm\mid l'm')$.  A similar
convention will be used for three-dimensional column vectors
representing quantities with a single index $\sigma$ (such as the
induced-force multipolar amplitudes).  With this notation, equation
\refeq{induced force equations for matrix elements} can be written in
the form
\begin{equation}
\label{induced force equations in matrix notation}
   \sum_{j=1}^N\sum_{l'm'}
      \GrandMobility_{ij}(lm\mid l'm')
   \bcdot
      \inducedForceMultipole_j(l'm')
      =\externalVelocityCoefficient_i(lm),
\end{equation}
with the matrix $\GrandMobility$ given by the relation
\begin{equation}
\label{Grand Mobility matrix}
      \GrandMobility_{ij}(lm\mid l'm')
   =
      \delta_{ij}\delta_{ll'}\delta_{mm'}\Zsingle_i^{-1}(l)
   +
      \GreenFree_{ij}(lm\mid l'm')
   +
      \GreenWall_{ij}(lm\mid l'm'),
\end{equation}
according to expressions \refeq{Grand Mobility matrix elements} and
\refeq{Z diagonal}.  In the above equations,
$\inducedForceMultipole_i(lm)$ and
$\externalVelocityCoefficient_i(lm)$ are column vectors with
components $f_i(lm\sigma)$ and $c_i(lm\sigma)$, and the dot represents
the matrix multiplication.  

An analogous matrix notation will be used in the Cartesian
representation, which is introduced in the following section.

\section{Cartesian basis}
\label{Cartesian basis}

Using the force-multipole equations \refeq{induced force equations in
matrix notation} to determine the hydrodynamic friction matrix in a
suspension bounded by planar walls involves evaluation of the
spherical matrix elements \refeq{w projections of perturbation Greens
tensor} of the Green function $\bT'(\br,\br')$ that describes the flow
field in the bounded domain.  For a single wall the matrix elements
were calculated by Cichocki
\etal~\cite{Cichocki-Jones:1998,Cichocki-Jones-Kutteh-Wajnryb:2000}
using a multipolar-image representation of the flow reflected from a
planar boundary.  For a suspension confined between two parallel
walls, the matrix elements \refeq{w projections of perturbation Greens
tensor} can be evaluated using the image representation derived by
Bhattacharya and Blawzdziewicz \cite{Bhattacharya-Blawzdziewicz:2002};
however such calculations are inefficient due to convergence problems.
Here we propose an alternative approach, in which the matrix elements
\refeq{induced force equations in matrix notation} are determined by
means of Cartesian representation of the flow fields, consistent with
the wall geometry.

In what follows we assume that the walls are normal to the axis $z$ in
the Cartesian coordinate system $(x,y,z)$.  The corresponding
Cartesian unit vectors are denoted $\ex,\ey,\ez$.

\subsection{Definition of Cartesian basis flow fields}
\label{Definition of basis flow fields}

We introduce two basis sets of Stokes flows
$\CartesianBasisM{\bk\sigma}$ and $\CartesianBasisP{\bk\sigma}$,
defined by the expressions
\begin{subequations}
\label{Cartesian basis -}
\begin{eqnarray}
\label{Cartesian basis - 0}
\CartesianBasisM{\bk0}(\br)&=&(32\upi^2)^{-1/2}\,
\left[
   \im(1-2kz)\hat\bk+(1+2kz)\ez
\right]
   k^{-1/2}\e^{\im\bk\bcdot\brho-kz},
\\\nonumber\\
\label{Cartesian basis - 1}
\CartesianBasisM{\bk1}(\br)&=&(8\upi^2)^{-1/2}\,
(
   \hat\bk\boldsymbol{\times}\ez
)
   k^{-1/2}\e^{\im\bk\bcdot\brho-kz},
\\\nonumber\\
\label{Cartesian basis - 2}
\CartesianBasisM{\bk2}(\br)&=&(32\upi^2)^{-1/2}\,
(
   \im\hat\bk-\ez
)
   k^{-1/2}\e^{\im\bk\bcdot\brho-kz},
\end{eqnarray}
\end{subequations}
and
\begin{subequations}
\label{Cartesian basis +}
\begin{eqnarray}
\label{Cartesian basis + 0}
\CartesianBasisP{\bk0}(\br)&=&(32\upi^2)^{-1/2}\,
(
   \im\hat\bk+\ez
)
   k^{-1/2}\e^{\im\bk\bcdot\brho+kz},
\\\nonumber\\
\label{Cartesian basis + 1}
\CartesianBasisP{\bk1}(\br)&=&(8\upi^2)^{-1/2}\,
(
\hat\bk\boldsymbol{\times}\ez
)
   k^{-1/2}\e^{\im\bk\bcdot\brho+kz},
\\\nonumber\\
\label{Cartesian basis + 2}
\CartesianBasisP{\bk2}(\br)&=&(32\upi^2)^{-1/2}\,
\left[
   \im(1+2kz)\hat\bk-(1-2kz)\ez
\right]
   k^{-1/2}\e^{\im\bk\bcdot\brho+kz}.
\end{eqnarray}
\end{subequations}
The pressure fields corresponding to the flows
\refeq{Cartesian basis -} and \refeq{Cartesian basis +} are
\begin{equation}
\label{Cartesian pressure -}
   p^-_{\bk0}(\br)=(2\upi^2)^{-1/2}\eta\,
   k^{1/2}\e^{\im\bk\bcdot\brho-kz},\qquad
   p^+_{\bk2}(\br)=(2\upi^2)^{-1/2}\eta\,
   k^{1/2}\e^{\im\bk\bcdot\brho+kz},
\end{equation}
and
\begin{equation}
\label{Cartesian pressure +}
   p^-_{\bk1}(\br)=p^-_{\bk2}(\br)=
   p^+_{\bk0}(\br)=p^+_{\bk1}(\br)=0.
\end{equation}
In the above relations
\begin{equation}
\label{rho}
\brho=x\ex+y\ey
\end{equation}
is the projection of the vector $\br$ onto the $x$--$y$ plane, and
\begin{equation}
\label{wave vector}
\bk=k_x\ex+k_y\ey
\end{equation}
is the corresponding two-dimensional wave vector.  Furthermore,
$\hat\bk=\bk/k$ and $k=|\bk|$.  

The basis sets \refeq{Cartesian basis -} and \refeq{Cartesian basis +}
are determined by the following conditions: Firstly, each basis flow
field $\CartesianBasisPM{\bk\sigma}$ corresponds to a single lateral
Fourier mode; secondly, the velocity fields
$\CartesianBasisM{\bk\sigma}(\br)$ vanish for $z\to\infty$ and
$\CartesianBasisP{\bk\sigma}(\br)$ for $z\to-\infty$; thirdly, the
basis fields $\CartesianBasisP{\bk\sigma}$ are obtained from
$\CartesianBasisM{\bk\,2-\sigma}$ by reflection with respect to the
plane $x$--$y$.  The fourth condition is the set of curl relations
\begin{subequations}
\label{Cartesian curl -}
\begin{equation}
\label{Cartesian curl - 1}
\CartesianBasisM{\bk1}
   =-\half\im k^{-1}
      \boldsymbol{\nabla}\boldsymbol{\times}\CartesianBasisM{\bk0},
\end{equation}
\begin{equation}
\label{Cartesian curl - 2}
\CartesianBasisM{\bk2}
   =-\half\im k^{-1}
      \boldsymbol{\nabla}\boldsymbol{\times}\CartesianBasisM{\bk1},
\end{equation}
and
\end{subequations}
\begin{subequations}
\label{Cartesian curl +}
\begin{equation}
\label{Cartesian curl + 1}
\CartesianBasisP{\bk1}
   =\half\im k^{-1}
      \boldsymbol{\nabla}\boldsymbol{\times}\CartesianBasisP{\bk2},
\end{equation}
\begin{equation}
\label{Cartesian curl + 0}
\CartesianBasisP{\bk0}
   =\half\im k^{-1}
      \boldsymbol{\nabla}\boldsymbol{\times}\CartesianBasisP{\bk1},
\end{equation}
\end{subequations}
by analogy to the expressions \refeq{spherical curl
-}--\refeq{spherical zero curl} for the spherical basis.  Relations
\refeq{Cartesian curl -} and \refeq{Cartesian curl +} imply
\refstepcounter{equation}
$$
\label{Cartesian zero curl}
   \bnabla\btimes\CartesianBasisM{\bk2}=0,
\qquad
   \bnabla\btimes\CartesianBasisP{\bk0}=0.
\eqno{(\theequation{\mathit{a},\mathit{b}})}
$$ 
The final condition is the requirement that the basis fields
\refeq{Cartesian basis -} and \refeq{Cartesian basis +} satisfy the
identity \refstepcounter{equation}
$$
\label{expansion of Oseen tensor in Cartesian basis}
\eta\bT_0(\br-\br')=
\left\{
   \begin{array}{ll}
     \displaystyle\int_{\bk}\diff\bk\sum_{\sigma}
        \CartesianBasisM{\bk\sigma}(\br)
        \CartesianBasisPcon{\bk\sigma}(\br'),
        \qquad &z>z',\\&\\
     \displaystyle\int_\bk\diff\bk\sum_{\sigma}
        \CartesianBasisP{\bk\sigma}(\br)
        \CartesianBasisMcon{\bk\sigma}(\br'),
        \qquad &z<z',
   \end{array}
\right.
\eqno{(\theequation{\mathit{a},\mathit{b}})}
$$ 
where the integration is over the two-dimensional Fourier space
\refeq{wave vector}.  The identity \refeq{expansion of Oseen tensor in
Cartesian basis} is analogous to the representation \refeq{expansion
of Oseen tensor in spherical basis} of the Oseen tensor in the
spherical basis.  It can be verified by showing that
\begin{equation}
\label{integrand of representation of Oseen tensor in Cartesian basis}
\CartesianBasisMP{\bk\sigma}(\br)
   \CartesianBasisPMcon{\bk\sigma}(\br')
      =\eta\hat\bT_0(\bk,z-z')\,\e^{\im\bk\bcdot(\brho-\brho')},
\end{equation}
where 
\begin{equation}
\label{xy Fourier transform of Oseen tensor}
\eta\hat\bT_0(\bk,z)
   =\frac{1}{16\upi^2}
      [2\identityTensor-(1+k|z|)\hat\bk\hat\bk
         -\im kz(\hat\bk\ez+\ez\hat\bk)
         -(1-k|z|)\ez\ez]k^{-1}\e^{-k|z|},
\end{equation}
is the two-dimensional Fourier transform in the $x$--$y$ plane of the
Oseen tensor, 
\begin{equation}
\label{definition of Fourier transform of Oseen tensor}
\hat\bT_0(\bk,z)
   =\frac{1}{(2\upi)^2}\int\bT_0(\br)\e^{-\im\bk\bcdot\brho}\diff\brho.
\end{equation}

The reciprocal sets of vector fields $\reciprocalCartesianBasisPM{\bk
\sigma}$ that correspond to the Cartesian basis sets
$\CartesianBasisPM{\bk \sigma}$ are defined by the orthogonality
relations
\begin{equation}
\label{orthogonality relations for Cartesian reciprocal basis}
\langle\deltah{h}\reciprocalCartesianBasisPM{\bk\sigma}\mid
    \CartesianBasisPM{\bk'\sigma'}\rangle
    =\delta(\bk-\bk')\delta_{\sigma\sigma'},
\end{equation}
which hold for all values of the parameter $h$, where
\begin{equation}
\label{Cartesian delta}
\deltah{h}(\br)=\delta(z-h).
\end{equation}
Inserting expressions \refeq{Cartesian basis -} and \refeq{Cartesian
basis +} into the above relations yields
\begin{equation}
\label{Cartesian reciprocal basis }
\reciprocalCartesianBasisPM{\bk0}(\br)
=4k\CartesianBasisMP{\bk0}(\br),
\qquad
\reciprocalCartesianBasisPM{\bk1}(\br)
=2k\CartesianBasisMP{\bk1}(\br),
\qquad
\reciprocalCartesianBasisPM{\bk2}(\br)
=4k\CartesianBasisMP{\bk2}(\br).
\end{equation}

\subsection{Displacement theorem for Cartesian basis}
\label{Displacement theorem for Cartesian basis}

The Cartesian basis fields \refeq{Cartesian basis -} and
\refeq{Cartesian basis +} centered at different points $\bR_1$ and
$\bR_2$ are related by simple displacement transformations.  Due to
the translational invariance, the transformations are diagonal in the
wave vector $\bk$,
\begin{subequations}
\label{Cartesian displacement theorems -+}
\begin{eqnarray}
\label{Cartesian displacement theorem -}
\CartesianBasisM{\bk\sigma}(\br_2)&=&\sum_{\sigma'}
   \CartesianBasisM{\bk\sigma'}(\br_1)
   \CartesianDisplacementElements{--}(\bR_{12},\bk;\sigma'\mid\sigma),
\\
\label{Cartesian displacement theorem +}
\CartesianBasisP{\bk\sigma}(\br_2)&=&\sum_{\sigma'}
   \CartesianBasisP{\bk\sigma'}(\br_1)
   \CartesianDisplacementElements{++}(\bR_{12},\bk;\sigma'\mid\sigma),
\end{eqnarray}
\end{subequations}
where $\br_1=\br-\bR_1$, $\br_2=\br-\bR_2$, and
$\bR_{12}=\bR_1-\bR_2$.  Using the orthogonality condition
\refeq{orthogonality relations for Cartesian reciprocal basis} and the
completeness of the Cartesian basis sets we get
\begin{equation}
\label{Cartesian displacement matrix in terms of reciprocal basis}
   \langle\deltah{0}(1)\reciprocalCartesianBasisPM{\bk'\sigma'}(1)
\mid
   \CartesianBasisPM{\bk\sigma}(2)\rangle
=\delta(\bk-\bk')
   \CartesianDisplacementElements{\pm\pm}(\bR_{12},\bk;\sigma'\mid\sigma).
\end{equation}
Relations \refeq{Cartesian displacement theorems -+} and the
expression \refeq{Cartesian displacement matrix in terms of reciprocal
basis} for the elements
$\CartesianDisplacementElements{\pm\pm}(\bR_{12},\bk;\sigma'\mid\sigma)$
of the Cartesian displacement matrix
$\CartesianDisplacement{\pm\pm}(\bR_{12},\bk)$ are analogous to the
displacement formulas \refeq{spherical displacement expression for
Oseen matrix elements}--\refeq{relation of Oseen matrix to spherical
displacement} for the spherical basis fields.

An analysis of equations \refeq{Cartesian basis -} and
\refeq{Cartesian basis +} indicates that the matrices
$\CartesianDisplacement{\pm\pm}(\bR_{12},\bk)$ can be written in the
factorized form
\begin{equation}
\label{factorization of Cartesian displacement matrix}
\CartesianDisplacement{\pm\pm}(\bR_{12},\bk)
   =\tildeCartesianDisplacement{\pm\pm}(kZ_{12})\e^{\im\bk\bcdot\brho_{12}},
\end{equation}
where
\begin{equation}
\label{definition of rho_12}
\bR_{12}=\brho_{12}+Z_{12}\ez,
\end{equation}
and
\begin{equation}
\label{expression for tilde S}
\tildeCartesianDisplacement{--}(kZ)=
\left[
   \begin{array}{ccc}
      1&0&0\\
      0&1&0\\
      -2kZ&0&1
   \end{array}
\right]
\be^{-kZ},\qquad
\tildeCartesianDisplacement{++}(kZ)=
\left[
   \begin{array}{ccc}
      1&0&2kZ\\
      0&1&0\\
      0&0&1
   \end{array}
\right]\be^{kZ}.
\end{equation}
It is also easy to verify that the matrices
$\CartesianDisplacement{\pm\pm}$ obey the group property
\begin{equation}
\label{group property of Cartesian displacement matrices}
\CartesianDisplacement{\pm\pm}(\bR+\bR',\bk)=
   \CartesianDisplacement{\pm\pm}(\bR,\bk)\bcdot
   \CartesianDisplacement{\pm\pm}(\bR',\bk)
\end{equation}
with
\begin{equation}
\label{unity of group of Cartesian displacement matrices}
\CartesianDisplacement{\pm\pm}(0,\bk)=\identityTensor,
\end{equation}
and that they satisfy the symmetry relation 
\begin{equation}
\label{symmetry of Cartesian displacement matrices}
\CartesianDisplacement{++}(\bR,\bk)
   =[\CartesianDisplacement{--}(-\bR,\bk)]^\dagger,
\end{equation}
where the dagger denotes the Hermitian conjugate.

\subsection{Transformations between Cartesian and spherical basis sets}
\label{Transformations between Cartesian and spherical basis sets}

One of the key results of our study is a set of transformation
relations between the spherical basis fields \refeq{spherical basis v
-} and \refeq{spherical basis v +} and the Cartesian basis fields
\refeq{Cartesian basis -} and \refeq{Cartesian basis +}.  We focus on
the transformations
\begin{equation}
\label{spherical - in Cartesian}
\sphericalBasisM{lm\sigma}(\br_2)=\int\diff\bk'\sum_{\sigma'}
   \CartesianBasisPM{\bk'\sigma'}(\br_1)
   \langle\deltah{0}(1)\reciprocalCartesianBasisPM{\bk'\sigma'}(1)
\mid
   \sphericalBasisM{lm\sigma}(2)\rangle
\end{equation}
and
\begin{equation}
\label{Cartesian  in spherical +}
\CartesianBasisPM{\bk\sigma}(\br_2)=\sum_{l'm'\sigma'}
   \sphericalBasisP{l'm'\sigma'}(\br_1)
   \langle\deltab{a}(1)\reciprocalSphericalBasisP{l'm'\sigma'}(1)
\mid
   \CartesianBasisPM{\bk\sigma}(2)\rangle
\end{equation}
that are needed for evaluating the spherical matrix elements
\refeq{w projections of perturbation Greens tensor} of the Green
function representing the flow scattered from the walls.

Recalling notation \refeq{definition of r_i} and definition
\refeq{spherical basis v -} we note that the basis fields
$\sphericalBasisM{lm\sigma}(\br_2)$ are singular at $\br=\bR_2$.
Accordingly, the transformation formula \refeq{spherical - in
Cartesian} is valid for $\pm(\bR_2-\bR_1)\bcdot\ez>0$.  We also note
that the integral defining the matrix element on the right side of
\refeq{spherical - in Cartesian} is not absolutely convergent for
$l+\sigma\le2$, because of the slow convergence of the field
\refeq{spherical basis v -} at infinity; the principal-value
interpretation of the integral is employed in this case.

Using the Cartesian displacement relations \refeq{Cartesian
displacement theorems -+}, the matrix elements on the right side of
equations \refeq{spherical - in Cartesian} and \refeq{Cartesian in
spherical +} can be factorized into the products of the displacement
matrices \refeq{factorization of Cartesian displacement matrix} and
the position-independent matrices $\TransformationCS{\pm-}(\bk,lm)$
and $\TransformationSC{+\pm}(lm,\bk)$,
\begin{equation}
\label{factorization CS}
   \langle\deltah{0}(1)\reciprocalCartesianBasisPM{\bk\sigma}(1)
\mid
   \sphericalBasisM{l'm'\sigma'}(2)\rangle
=
\left[
   \CartesianDisplacement{\pm\pm}(\bR_{12},\bk)
   \bcdot\TransformationCS{\pm-}(\bk,l'm')
\right]_{\sigma\sigma'}\,,
\end{equation}
\begin{equation}
\label{factorization SC}
   \langle\deltab{a}(1)\reciprocalSphericalBasisP{lm\sigma}(1)
\mid
   \CartesianBasisPM{\bk'\sigma'}(2)\rangle
=
\left[
   \TransformationSC{+\pm}(lm,\bk')
   \bcdot\CartesianDisplacement{\pm\pm}(\bR_{12},\bk')
\right]_{\sigma\sigma'}\,.
\end{equation}
Equation \refeq{factorization of
Cartesian displacement matrix} indicates that the matrix element
\refeq{factorization CS} is nonsingular in the limit $\bR_{12}\to0$,
even though the integrand in the scalar product is singular at $\br=0$
for $\bR_1=\bR_2$.  (In contrast, relation \refeq{factorization SC}
does not involve any singular integrals.)  In the limit $\bR_{12}\to0$,
equations \refeq{unity of group of Cartesian displacement matrices}
and \refeq{spherical - in Cartesian}--\refeq{factorization SC} yield
the transformation relations
\begin{equation}
\label{spherical - in Cartesian -- same point}
\sphericalBasisM{lm\sigma}(\br)=\int\diff\bk'\sum_{\sigma'}
   \CartesianBasisPM{\bk'\sigma'}(\br)
      \TransformationElementCS{\pm-}(\bk',lm;\sigma'\mid\sigma),
\qquad \pm z<0,
\end{equation}
\begin{equation}
\label{Cartesian - in spherical -- same point}
\CartesianBasisPM{\bk\sigma}(\br)=\sum_{l'm'\sigma'}
   \sphericalBasisP{l'm'\sigma'}(\br)
      \TransformationElementSC{+\pm}(l'm',\bk;\sigma'\mid\sigma).
\end{equation}

The matrices $\TransformationCS{\pm-}(\bk,lm)$ and
$\TransformationSC{+\pm}(lm,\bk)$ have several important symmetries.
First, we recall that the Cartesian basis sets \refeq{Cartesian basis
-} and \refeq{Cartesian basis +} are related to each other via the
reflection with respect to the plane $z=0$. The corresponding
symmetries of the transformation matrices are
\begin{equation}
\label{reflection symmetry of Tcs}
\TransformationCS{--}(\bk,lm;\sigma\mid\sigma')
   =(-1)^{l+m+\sigma'}\TransformationCS{+-}(\bk,lm;2-\sigma\mid\sigma'),
\end{equation}
\begin{equation}
\label{reflection symmetry of Tsc}
\TransformationSC{+-}(lm,\bk;\sigma\mid\sigma')
   =(-1)^{l+m+\sigma}\TransformationSC{++}(lm,\bk;\sigma\mid2-\sigma').
\end{equation}

Another important symmetry relation is associated with the
representations \refeq{expansion of Oseen tensor in spherical basis}
and \refeq{expansion of Oseen tensor in Cartesian basis} of the Oseen
tensor in the spherical and Cartesian bases.  The relation is obtained
by applying the Oseen integral operator with the kernel in the
respective forms (\ref{expansion of Oseen tensor in spherical
basis}\textit{b}) and \refeq{expansion of Oseen tensor in Cartesian
basis} to the fields $\reciprocalCartesianBasisPM{\bk\sigma}$, which
yields
\begin{equation}
\label{Oseen operator applied to Cartesian W -- spherical representation}
\int\bT_0(\br_1-\br'_1)
   \deltah{0}(\br'_2)\reciprocalCartesianBasisPM{\bk\sigma}(\br'_2)\diff\br'
    =\sum_{l'm'\sigma'}
      \sphericalBasisP{l'm'\sigma'}(\br_1)
      \langle\sphericalBasisM{l'm'\sigma'}(1)
   \mid
      \deltah{0}(2)\reciprocalCartesianBasisPM{\bk\sigma}(2)\rangle
\end{equation}
and
\begin{equation}
\label{Oseen operator applied to Cartesian W -- Cartesian representation}
\int\bT_0(\br_1-\br'_1)
   \deltah{0}(\br'_2)\reciprocalCartesianBasisPM{\bk\sigma}(\br'_2)\diff\br'
=\int\diff\bk'\sum_{\sigma'}
      \CartesianBasisMP{\bk'\sigma'}(\br_1)
      \langle\CartesianBasisPM{\bk'\sigma'}(1)
   \mid
      \deltah{0}(2)\reciprocalCartesianBasisPM{\bk\sigma}(2)\rangle.
\end{equation}
By comparing the above expressions in the limit $\bR_{12}\to0$ we find
\begin{equation}
\label{reciprocal expansion for Cartesian v in spherical basis}
\CartesianBasisMP{\bk\sigma}(\br)
   =\sum_{l'm'\sigma'}
      \sphericalBasisP{l'm'\sigma'}(\br)
      \TransformationElementCScon{\pm-}(\bk,lm;\sigma\mid\sigma'),
\end{equation}
where equations \refeq{unity of group of Cartesian displacement
matrices} and \refeq{factorization CS} and the orthogonality
condition \refeq{orthogonality relations for Cartesian reciprocal
basis} were applied.  Since the expansion of the flow fields
$\CartesianBasisPM{\bk\sigma}(\br)$ into
$\sphericalBasisP{l'm'\sigma'}(\br)$ is unique, equations
\refeq{Cartesian - in spherical -- same point} and \refeq{reciprocal
expansion for Cartesian v in spherical basis} imply the symmetry
\begin{equation}
\label{CS--SC symmetry}
\TransformationSC{+\mp}(lm,\bk)
  =\left[\TransformationCS{\pm-}(\bk,lm)\right]^\dagger.
\end{equation}

The functional dependence of the matrices $\TransformationSC{+\pm}$
and $\TransformationCS{\pm-}$ on the wave vector $\bk$ can also be
derived using symmetry considerations.  Specifically, one can show
that
\begin{subequations}
\label{functional form of SC and CS transformations  T on k}
\begin{eqnarray}
\label{functional form of SC transformation  T on k}
   &&\TransformationElementSC{+\pm}(lm,\bk;\sigma\mid\sigma')
      \sim k^{-1/2}k^{l+\sigma-1}\e^{\im m\psi},
\\
\label{functional form of CS transformation  T on k}
   &&\TransformationElementCS{\pm-}(\bk,lm;\sigma\mid\sigma')
      \sim k^{-1/2}k^{l+\sigma'-1}\e^{-\im m\psi},
\end{eqnarray}
\end{subequations}
where $\bk=(k,\psi)$ is the representation of the wave vector in the
polar coordinates.  The angular form of relations \refeq{functional
form of SC and CS transformations T on k} stems from the requirement
in the definition of the basis fields \refeq{spherical basis v -} and
\refeq{spherical basis v +} that the coefficients
$\sphericalBasisCoefPM{lm\sigma}(\theta,\phi)$ are combinations of
spherical harmonics of order $m$.  The dependence on the amplitude of
the wave vector $k$ results from the invariance of the transformation
relations \refeq{spherical - in Cartesian -- same point} and
\refeq{Cartesian - in spherical -- same point} with respect to the
scale change
\begin{equation}
\label{scale change}
\br\to\alpha\br,\qquad\bk\to\alpha^{-1}\bk,
\end{equation}
where $\alpha$ is a real parameter (cf., expressions \refeq{spherical
basis v -}, \refeq{spherical basis v +} and \refeq{Cartesian basis
-}, \refeq{Cartesian basis +} for the spherical and Cartesian basis
fields).  Using equations \refeq{functional form of SC and CS
transformations T on k}, the matrices
$\TransformationSC{+\pm}(lm,\bk)$ and
$\TransformationCS{\pm-}(\bk,lm)$ can be represented in the factorized
form
\begin{subequations}
\label{factorization of transformations SC and CS}
\begin{equation}
\label{factorization of transformation SC}
\TransformationSC{+\pm}(lm,\bk)
   =(-\im)^m(2\upi k)^{-1/2}\e^{-\im m\psi}
      \Kmatrix(k,l)\bcdot\tildeTransformationSC{+\pm}(lm),
\end{equation}
\begin{equation}
\label{factorization of transformation CS}
\TransformationCS{\pm-}(\bk,lm)
   =\im^m(2\upi k)^{-1/2}\e^{\im m\psi}
      \tildeTransformationCS{\pm-}(lm)\bcdot\Kmatrix(k,l),
\end{equation}
\end{subequations}
where $\Kmatrix(k,l)$ is a diagonal matrix with the elements
\begin{equation}
\label{matrix K}
\KmatrixElement(k,l;\sigma\mid\sigma')=\delta_{\sigma\sigma'}k^{l+\sigma-1}.
\end{equation}

A further simplification of the structure of the transformation
matrices $\TransformationSC{+\pm}$ and $\TransformationCS{\pm-}$
results from the curl relations \refeq{spherical curl
-}--\refeq{spherical zero curl} and \refeq{Cartesian curl
-}--\refeq{Cartesian zero curl}.  By taking curl of both sides of
equations \refeq{spherical - in Cartesian -- same point} and
\refeq{Cartesian - in spherical -- same point}, applying the
symmetries \refeq{reflection symmetry of Tcs}, \refeq{reflection
symmetry of Tsc}, and \refeq{CS--SC symmetry}, and using the
factorization formulas \refeq{factorization of transformations SC and
CS}, one can show that the matrices
$\tildeTransformationSC{+\pm}(lm)$ and
$\tildeTransformationCS{\pm-}(lm)$ have the following triangular
structure
\begin{subequations}
\label{form of transformation SC and CS}
\begin{equation}
\label{form of transformation SC}
   \tildeTransformationSC{++}=
\left[
\begin{array}{ccc}
      a&b&c\\
      0&2a&2b\\
      0&0&4a
\end{array}
\right],
\qquad
   \tildeTransformationSC{+-}=
(-1)^{l+m}\left[
\begin{array}{ccc}
      c&b&a\\
      -2b&-2a&0\\
      4a&0&0
\end{array}
\right],
\end{equation}
\begin{equation}
\label{form of transformation CS}
   \tildeTransformationCS{+-}=
(-1)^{l+m}\left[
\begin{array}{ccc}
      c&-2b&4a\\
      b&-2a&0\\
      a&0&0
\end{array}
\right],
\qquad
   \tildeTransformationCS{--}=
\left[
\begin{array}{ccc}
      a&0&0\\
      b&2a&0\\
      c&2b&4a
\end{array}
\right],
\end{equation}
\end{subequations}
and involve only three independent coefficients.  As shown in Appendix
\ref{Derivation of coefficients a,b,c}, the expressions for the
coefficients $a,b,c$ are
\begin{subequations}
\label{coefficients a,b,c}
\begin{eqnarray}
\label{coefficient a}
a&=&[4(l-m)!(l+m)!(2l+1)]^{-1/2},\\\nonumber\\
\label{coefficient b}
b&=&2am/l,\\\nonumber\\
\label{coefficient c}
c&=&a\frac{l(2l^2-2l-1)-2m^2(l-2)}{l(2l-1)}.
\end{eqnarray}
\end{subequations}

Relations \refeq{factorization of Cartesian displacement
matrix}--\refeq{symmetry of Cartesian displacement matrices} and
\refeq{factorization of transformations SC and
CS}--\refeq{coefficients a,b,c} represent the key results of the
analysis presented so far.  In Section \ref{Fourier representation of
displacement matrix in spherical basis} we apply these results to
express the spherical matrix elements of the free-space Green operator
\refeq{w projections of Oseen tensor} in terms of two-dimensional
Fourier integrals (which can be explicitly evaluated in this case).
The Fourier representations of matrix elements \refeq{w projections of
perturbation Greens tensor} for a system bounded by a single planar
wall and by two parallel planar walls are derived in \S
\ref{Hyrodynamic interactions with single planar wall} and \S
\ref{Suspension between two planar walls--matrix elements},
respectively.  These results enable efficient numerical evaluation of
the multiparticle friction matrix in wall-bounded suspensions.
Description of our algorithm is given in Section \ref{Numerical
implementation}, and examples of numerical results are
shown in Section \ref{Results}.

\section{Fourier representation of the spherical displacement matrix}
\label{Fourier representation of displacement matrix in spherical basis}

In this section we apply the Cartesian displacement formulas
\refeq{expression for tilde S} and transformation matrices
\refeq{factorization of transformations SC and CS}--\refeq{form of
transformation SC and CS} to express the spherical displacement matrix
$\sphericalDisplacement{+-}$ in terms of two-dimensional Fourier
integrals.  Such a representation can be utilized in developing
numerical algorithms for evaluating hydrodynamic interactions in
doubly periodic systems.  Moreover, the analysis allows us to
introduce some techniques that will be used in the discussion of the
flow in wall-bounded suspensions (cf.\ Sections \ref{Hyrodynamic
interactions with single planar wall} and \ref{Suspension between
two planar walls--matrix elements}).

We recall that the displacement matrix $\sphericalDisplacement{+-}$
and the corresponding spherical matrix elements \refeq{w projections
of Oseen tensor} of the Oseen operator are equivalent, as indicated by
the formula \refeq{relation of Oseen matrix to spherical
displacement}.  To make the notation in this and the following
sections parallel, we express our results in terms of the matrix
elements $\GreenFree_{ij}(lm\mid l'm')$.  

By inserting the expansion \refeq{spherical - in Cartesian -- same
point} into \refeq{spherical displacement expression for Oseen matrix
elements} and using equation \refeq{factorization SC} we find
\begin{equation}
\label{Fourier representation of spherical displacement - Green notation}
   \GreenFree_{ij}(lm\mid l'm')=
\eta^{-1}\int\diff\bk\,
   \TransformationSC{+\pm}(lm,\bk)
\bcdot
   \CartesianDisplacement{\pm\pm}(\bR_{ij},\bk)
\bcdot
   \TransformationCS{\pm-}(\bk,l'm'),
\end{equation}
where the plus sign applies for $\bR_{ij}\bcdot\ez<0$ and the minus
sign for $\bR_{ij}\bcdot\ez>0$.  We note that the Lorentz symmetry
\refeq{reciprocal relations elements of Oseen operator} of the matrix
elements \mbox{$\GreenFree_{ij}(lm\mid l'm')$} is explicit in equation
\refeq{Fourier representation of spherical displacement - Green
notation} due to the symmetry relations \refeq{symmetry of Cartesian
displacement matrices} and \refeq{CS--SC symmetry} for the component
matrices.  The angular integration in relation \refeq{Fourier
representation of spherical displacement - Green notation} can be
explicitly performed with a help of the factorization formulas
\refeq{factorization of Cartesian displacement matrix} and
\refeq{factorization of transformations SC and CS} and the equation
\begin{equation}
\label{angular integral of Fourier mode}
\int_0^{2\upi}\e^{\im (k\rho\cos\psi-m\psi)}d\psi=2\upi\im^m \BesselJ_m(k\rho),
\end{equation}
where $\BesselJ_m(x)$ is the Bessel function of the order $m$.  The
resulting expression has the form
\begin{eqnarray}
\label{Fourier representation of spherical displacement factorized}
   &&\GreenFreeElement_{ij}(lm\sigma\mid l'm'\sigma')=
   \eta^{-1}(-1)^{m'-m}
   \e^{\im(m'-m)\phi_{ij}}
\nonumber \\&&\hphantom{\GreenFreeElement_{ij}(lm\sigma}
   \times
      \int_0^{\infty}
      \FreeGreenFourierElementPM(kZ_{ij};lm\sigma\mid l'm'\sigma')
         k^{l+l'+\sigma+\sigma'-2}\BesselJ_{m-m'}(k\rho_{ij})\diff k,
\end{eqnarray}
where $(\rho_{ij},\phi_{ij},Z_{ij})$ is the representation of the
vector $\bR_{ij}$ in the cylindrical coordinates, and
\begin{equation}
\label{expression for g_0}
\FreeGreenFourierPM(kz;lm\mid l'm')=
   \tildeTransformationSC{+\pm}(lm)
\bcdot
   \tildeCartesianDisplacement{\pm\pm}(kz)
\bcdot
   \tildeTransformationCS{\pm-}(l'm').
\end{equation}

Relations \refeq{expression for tilde S} and \refeq{expression for
g_0} indicate that the integrand in equation \refeq{Fourier
representation of spherical displacement factorized} is a combination
of the Bessel function, powers of $k$, and the exponential
$\e^{-k|Z_{ij}|}$.  The integrals of this form can be evaluated using
the following identity
\begin{equation}
\label{Sukalyan's identity for integral of Bessel functions}
\int_0^\infty k^l\BesselJ_m(k\rho)\e^{-kz}\diff k
   =(-1)^m(l-m)!\,r^{-(l+1)}\LegendreP_l^m(r^{-1}z),\quad z>0,
\end{equation}
where $P_l^m(x)$ is the associated Legendre polynomial, and
\begin{equation}
\label{radius in Legendre polynomial}
r=(\rho^2+z^2)^{1/2}.
\end{equation}
We have verified that equations \refeq{Fourier representation of
spherical displacement factorized}--\refeq{Sukalyan's identity for
integral of Bessel functions} yield expressions that are equivalent to
the displacement theorems for the spherical basis of Stokes flows
derived by Felderhof and Jones \cite{Felderhof-Jones:1989}.  In
development of the multipolar-expansion algorithms to evaluate
hydrodynamic interactions in doubly-periodic systems, a direct
application of the Fourier representation \refeq{Fourier
representation of spherical displacement - Green notation} may be
useful.

\section{Single-wall Green operator}
\label{Hyrodynamic interactions with single planar wall}

Similar techniques can be used to evaluate the matrix elements of the
Green operator \refeq{w projections of perturbation Greens tensor} in
a system bounded by a single planar wall.  We assume that the wall is
in the plane
\begin{equation}
\label{position of wall}
z=\Zwall{}
\end{equation}
and the suspension occupies either the halfspace $z>\Zwall{}$ (denoted
by $\halfspace{+}$) or $z<\Zwall{}$ (denoted by $\halfspace{-}$).  The
spherical matrix elements of the Green operator \refeq{w projections
of perturbation Greens tensor} for this system are obtained by
combining the transformation relations \refeq{factorization of
transformations SC and CS}--\refeq{form of transformation SC and CS}
between the spherical and Cartesian basis sets with the Cartesian
representation of the flow reflected from the wall.  The reflected
flow is discussed in the following subsection.

\subsection{Single-wall reflection matrix}
\label{Single-wall reflection matrix}

The velocity field in the halfspace $\halfspace{\pm}$, occupied by the
fluid, can be uniquely decomposed into the incoming and reflected
flows
\begin{equation}
\label{decomposition of flow into incoming and reflected components}
\bv(\br)=\vWin(\br)+\vWout(\br),
\end{equation}
where
\begin{subequations}
\label{incoming and reflected flow for one wall}
\begin{equation}
\label{incoming flow}
\vWin(\br)
  =\int\diff\bk\sum_\sigma\cWinCoef{\bk\sigma}
                     \CartesianBasisPM{\bk\sigma}(\rwall{}),
\end{equation}
\begin{equation}
\label{outcoming flow}
\vWout(\br)
  =\int\diff\bk\sum_\sigma \cWoutCoef{\bk\sigma}
                     \CartesianBasisMP{\bk\sigma}(\rwall{}).
\end{equation}
\end{subequations}
In the above relations 
\begin{equation}
\label{r wall}
\rwall{}=\br-\Rwall{}
\end{equation}
denotes the position of the point $\br$ relative to the wall, where
$\Rwall{}=(\Xwall{},\Ywall{},\Zwall{})$ has arbitrary lateral
coordinates $\Xwall{}$ and $\Ywall{}$.  

According to the definitions \refeq{Cartesian basis -} and
\refeq{Cartesian basis +}, the decay of the basis flow fields
$\CartesianBasisMP{\bk\sigma}(\rwall{})$ for $k\to\infty$ is faster in
the halfspace $\halfspace{\pm}$ than it is on the wall surface
\refeq{position of wall}.  Thus, assuming that the integral
\refeq{outcoming flow} converges on the surface \refeq{position of
wall}, the scattered flow field $\vWout(\br)$ is nonsingular in the
whole region $\halfspace{\pm}$ occupied by the fluid.  Likewise, the
convergence of the integral \refeq{incoming flow} on the wall surface
implies that the incoming flow field $\vWin(\br)$ is nonsingular in
the complementary region $\halfspace{\mp}$.

By analogy with the relations \refeq{one-particle induced force
equation} and \refeq{multipolar expansion of scattered field} for a
flow field scattered by a particle, we introduce the single-wall
scattering matrix $\ZsingleWall$, defined by the equation
\begin{equation}
\label{definition of Z wall}
\cWout(\bk)=-\ZsingleWall
   \bcdot\cWin(\bk),
\end{equation}
where $\cWout(\bk)$ and $\cWin(\bk)$ denote the arrays of expansion
coefficients in equations \refeq{incoming and reflected flow for one
wall}.  For an immobile rigid wall, the velocity field
\refeq{decomposition of flow into incoming and reflected components}
vanishes at $z=\Zwall{}$.  By inspection of expressions
\refeq{Cartesian basis -} and \refeq{Cartesian basis +} we find that
\begin{equation}
\label{expression for Z wall}
\ZsingleWall
   =\left[
       \begin{array}{ccc}
          1&0&0\\
          0&1&0\\
          0&0&1
       \end{array}
    \right]
\end{equation}
in this case.  For planar interfaces with other boundary conditions
(e.g., a surfactant-covered fluid-fluid interface discussed in
\cite{Blawzdziewicz-Cristini-Loewenberg:1999}) the scattering matrix
is different from identity, and it may depend on the magnitude of the
wave vector $k$.  Explicit expressions for scattering matrices for
such systems will be derived in forthcoming publications.

\subsection{Matrix elements of Green operator}
\label{Matrix elements of Green operator - single wall}

In order to evaluate matrix elements \refeq{w projections of
perturbation Greens tensor} of the single-wall Green operator we
consider the flow field produced by the induced-force distribution
\refeq{induced forces} centered at the position of particle $j$.  By
comparing the decompositions \refeq{Green's function} and
\refeq{decomposition of flow into incoming and reflected components}
and using relation \refeq{spherical - in Cartesian -- same point} we
find
\begin{equation}
\label{incoming flow produced by force---one wall}
\vWin(\br)
=\int\bT_0(\br-\br')\bcdot\bF_j(\br')\diff\br'
\end{equation}
and
\begin{equation}
\label{outcoming flow produced by force---one wall}
\vWout(\br)=\int\bT'(\br,\br')\bcdot\bF_j(\br')\diff\br'.
\end{equation}
We note that according to equations \refeq{flow scattered by particle}
and \refeq{incoming flow produced by force---one wall} the flow
incident to the wall equals to the flow scattered by the particle
\begin{equation}
\label{incoming flow on wall equals outcoming from particle}
\vWin(\br)=\scatteredVelocity{j}(\br).
\end{equation}

By projecting equation \refeq{outcoming flow produced by force---one
wall} onto the reciprocal spherical basis
$\reciprocalSphericalBasisP{lm\sigma}$ centered at the position of
particle $i$ and using the multipolar expansion \refeq{induced force
multipoles} we get
\begin{equation}
\label{w projection of flow reflected from single wall}
\langle
   \deltab{a}(i)\reciprocalSphericalBasisP{lm\sigma}(i)
\mid
   \vWout\rangle
=\sum_{l'm'\sigma'}
   \GreenWallElement_{ij}(lm\sigma\mid l'm'\sigma')f_j(l'm'\sigma'),
\end{equation}
where the definition \refeq{w projections of perturbation Greens
tensor} was applied.  The matrix element of the reflected flow at the
left side of the above equation is evaluated with the help of the
reflection relation \refeq{definition of Z wall}.  Accordingly, the
expansion coefficients of the incoming flow
\begin{equation}
\label{expansion coefficients of incoming flow}
\cWinCoef{\bk\sigma}
   =\langle\deltah{0}(w)\reciprocalCartesianBasisPM{\bk\sigma}(w)
       \mid\vWin\rangle
\end{equation}
(where the index $w$ in the bra
$\langle\deltah{0}(w)\reciprocalCartesianBasisPM{\bk\sigma}(w)|$
indicates the dependence on the variable \refeq{r wall}) are
determined using expansion \refeq{multipolar expansion of scattered
field} for the incoming velocity field \refeq{incoming flow on wall
equals outcoming from particle} and the relation \refeq{factorization
CS} for the matrix elements relating the spherical and reciprocal
basis fields.  Collecting these formulas yields
\begin{equation}
\label{coefficents of incoming flow in terms of T and S matrices}
\cWin(\bk)=\eta^{-1}
   \sum_{l'm'}\CartesianDisplacement{\pm\pm}(\bR_{\wall j},\bk)
\bcdot
   \TransformationCS{\pm-}(\bk,l'm')
   \bcdot\inducedForceMultipole_j(l'm'),
\end{equation}
where $\bR_{i\wall}=\bR_i-\Rwall{}$ and $\bR_{\wall
j}=\Rwall{}-\bR_j$.  The above relation is combined with the expansion
\refeq{outcoming flow} and the scattering formula \refeq{definition of
Z wall}; the resulting expression for $\vWout$ is inserted into
\refeq{w projection of flow reflected from single wall}. The matrix
elements between the spherical and Cartesian basis fields are then
evaluated using \refeq{factorization SC}.  By comparing the result of
this calculation to the expression \refeq{w projection of flow
reflected from single wall} we find
\begin{equation}
\label{expression for single wall G'}
\GreenWall_{ij}(lm\mid l'm')=
-\eta^{-1}\int\diff\bk\,
   \TransformationSC{+\mp}(lm,\bk)
\bcdot
   \CartesianDisplacement{\mp\mp}(\bR_{i\wall},\bk)
\bcdot
\ZsingleWall
\bcdot
   \CartesianDisplacement{\pm\pm}(\bR_{\wall j},\bk)
\bcdot
   \TransformationCS{\pm-}(\bk,l'm').
\end{equation}

A physical interpretation of the above relation is straightforward:
the spherical components of the flow produced by the particle at point
$j$ are transformed into the Cartesian basis by the matrix
$\TransformationCS{\pm-}$; the Cartesian components are propagated by
the matrix $\CartesianDisplacement{\pm\pm}(\bR_{\wall j})$ to the
wall, where they are scattered (as represented by matrix
$\ZsingleWall$); the reflected field is propagated to the point $i$ by
the matrix $\CartesianDisplacement{\mp\mp}(\bR_{i\wall})$; and finally
the flow is transformed back into the spherical basis by the matrix
$\TransformationSC{+\mp}$.  We note that, similar to relation
\refeq{Fourier representation of spherical displacement - Green
notation}, the Lorentz symmetry \refeq{reciprocal relations elements
wall Green operator} of the matrix elements \refeq{expression for
single wall G'} is explicit due to the symmetry relations
\refeq{symmetry of Cartesian displacement matrices} and \refeq{CS--SC
symmetry} of the component matrices.

Similar to the angular integral in equation \refeq{Fourier
representation of spherical displacement - Green notation}, the
angular integration in the Fourier representation \refeq{expression
for single wall G'} of the matrix $\GreenWall_{ij}$ can be explicitly
performed with the help of expressions \refeq{factorization of
Cartesian displacement matrix}, \refeq{factorization of
transformations SC and CS}, and \refeq{angular integral of Fourier
mode}.  The resulting expression for the matrix elements of the
one-wall Green operator is
\begin{eqnarray}
\label{Fourier representation of single wall Green elements}
   &&\GreenWallElement_{ij}(lm\sigma\mid l'm'\sigma')=
   \eta^{-1}(-1)^{m'-m}
   \e^{\im(m'-m)\phi_{ij}}
\nonumber \\&&\hphantom{\GreenWallElement_{ij}}
   \times
      \int_0^{\infty}
      \SingleWallGreenFourierElementPM(
                          kZ_{i\wall},kZ_{\wall j};lm\sigma\mid l'm'\sigma')
         k^{l+l'+\sigma+\sigma'-2}\BesselJ_{m'-m}(k\rho_{ij})\diff k,
\end{eqnarray}
where
\begin{equation}
\label{expression for g single wall}
\SingleWallGreenFourierPM(kZ_{i\wall},kZ_{\wall j};lm\mid l'm')=
   -\tildeTransformationSC{+\mp}(lm)
\bcdot
   \tildeCartesianDisplacement{\mp\mp}(kZ_{i\wall})
\bcdot
   \ZsingleWall
\bcdot
   \tildeCartesianDisplacement{\pm\pm}(kZ_{\wall j})
\bcdot
   \tildeTransformationCS{\pm-}(l'm').
\end{equation}

We recall that the upper signs in the above equations correspond to
the fluid occupying the upper half-space $\halfspace{+}$, and the
lower signs to the fluid in the lower halfspace $\halfspace{-}$.
Taking this into account, we find that the exponential factors
resulting from the Cartesian displacement matrices \refeq{expression
for tilde S} in the product on the right side of equation
\refeq{expression for g single wall} can be combined into a single
exponential factor
\begin{equation}
\label{exponential factor for single wall}
\SingleWallGreenFourierPM(kZ_{i\wall},kZ_{\wall j};lm\mid l'm')
   \sim\e^{-k\imageOffset_{ij}}
\end{equation}
where
\begin{equation}
\label{offset - one wall}
\imageOffset_{ij}=|Z_{i\wall}|+|Z_{\wall j}|
\end{equation}
is the vertical offset between the target point $i$ at the position
$\bR_i$ and the image of the source point $j$ at
\begin{equation}
\label{position of image of target point}
\bR'_j=\bR_j-2(Z_j-\Zwall{})\ez.
\end{equation}
It follows that the integrand in equation \refeq{Fourier
representation of single wall Green elements} is the combination of
the factor \refeq{exponential factor for single wall}, the Bessel
function, and powers of $k$.  Thus, relations \refeq{Sukalyan's
identity for integral of Bessel functions} and \refeq{radius in
Legendre polynomial} imply that the elements of the matrix
$\SingleWallGreenFourierPM$ can be expressed in terms of the flow
produced by an image singularity at $\br=\bR'_j$.  We note that such a
form is required by the Lorentz reflection relation
\cite{Lorentz:1907}.  Explicit expressions for the image force
multipole system corresponding to an arbitrary source force multipole
have recently been derived by Cichocki and Jones
\cite{Cichocki-Jones:1998}; we have verified that the integral
\refeq{Fourier representation of single wall Green elements} yields
results equivalent to their expressions.

The main application of the Fourier representation of the single-wall
Green operator $\GreenWall_{ij}$ is in the subtraction technique that
is implemented in our algorithm for evaluating the multiparticle
friction matrix in a two-wall system.  In this application (in more
detail described in Section \ref{Integrands}) expressions
\refeq{Fourier representation of single wall Green elements} and
\refeq{expression for g single wall} are used in conjunction with the
results of Ref.~\cite{Cichocki-Jones:1998} to accelerate the
convergence of the Fourier integrals for the two-wall Green operator.

\section{Two-wall Green operator}
\label{Suspension between two planar walls--matrix elements}

In this section we generalize the analysis of Section \ref{Hyrodynamic
interactions with single planar wall} to a system of particles
confined between two parallel planar walls.  We assume that the walls
are in the planes
\begin{equation}
\label{positions of two walls}
z=\Zlow{}, \qquad z=\Zup{},
\end{equation}
where
\begin{equation}
\label{relative position of walls}
\Zlow{}<\Zup{}.
\end{equation}
The suspension occupies the region $\Zlow{}<z<\Zup{}$.  The positions
of the walls are indicated by vectors
$\Rlow{}=(\Xlow{},\Ylow{},\Zlow{})$ and
$\Rup{}=(\Xup{},\Yup{},\Zup{})$, where the lateral coordinates
$\Xlow{}$ and $\Ylow{}$ and $\Xup{}$ and $\Yup{}$ are chosen
arbitrarily.

The flow produced in this system by the induced-force distribution
\refeq{induced forces} centered at the position of particle $j$ is a
superposition of three components
\begin{equation}
\label{flow between two walls}
\bv(\br)=\vLout(\br)+\vUout(\br)+\scatteredVelocity{j}(\br).
\end{equation}
Here $\scatteredVelocity{j}(\br)$ is the velocity field \refeq{flow
scattered by particle} produced by force distribution $\bF_j$, and
$\vAout(\br)$ is the flow reflected by wall $\alpha=\low,\up$.  By
definition, the flow component $\vLout(\br)$ is nonsingular in the
region $z>\Zlow{}$ and vanishes for $z\to\infty$, and the flow
component $\vUout(\br)$ is nonsingular in the region $z<\Zup{}$ and
vanishes for $z\to-\infty$.  Accordingly, the expansions of the flow
fields $\vLout$ and $\vUout$ in the Cartesian basis sets
\refeq{Cartesian basis -} and \refeq{Cartesian basis +} have the form
\begin{subequations}
\label{expansion of v L and U out in Cartesian basis}
\begin{equation}
\label{expansion of v L out in Cartesian basis}
\vLout(\br)=\int\diff\bk\sum_\sigma\cLoutCoef{\bk\sigma}
                     \CartesianBasisM{\bk\sigma}(\rlow{}),
\end{equation}
\begin{equation}
\label{expansion of v U out in Cartesian basis}
\vUout(\br)=\int\diff\bk\sum_\sigma\cUoutCoef{\bk\sigma}
                     \CartesianBasisP{\bk\sigma}(\rup{}),
\end{equation}
\end{subequations}
where $\rlow{}=\br-\Rlow{}$ and $\rup{}=\br-\Rup{}$.  Expressions
\refeq{expansion of v L and U out in Cartesian basis} are consistent
with the expansion \refeq{outcoming flow}.

The three components \refeq{flow between two walls} of the velocity
field produced in the space between the walls by the force
distribution $\bF_j$ can be used to construct the flow components
$\vAin$ ($\alpha=\low,\up$) incoming to wall $\alpha$.  Using
expressions \refeq{expansion of v L and U out in Cartesian basis} and
the definition \refeq{incoming flow} of the incoming flow we find
\begin{subequations}
\label{incoming and outcoming flows for lower and upper walls}
\begin{equation}
\label{incoming and outcoming flows for lower wall}
   \vLin(\br)=\vUout(\br)+\scatteredVelocity{j}(\br),
\end{equation}
\begin{equation}
\label{incoming and outcoming flows for upper wall}
   \vUin(\br)=\vLout(\br)+\scatteredVelocity{j}(\br).
\end{equation}
\end{subequations}

Relation \refeq{flow scattered by particle} and the respective
decompositions \refeq{Green's function} and \refeq{flow between two
walls} of the Green function $\bT(\br,\br')$ and the flow field
$\bv(\br)$ imply that
\begin{equation}
\label{perturbation flow for two walls}
\vLout(\br)+\vUout(\br)=\int\bT'(\br,\br')\bcdot\bF_i(\br')\diff\br'.
\end{equation}
Projecting the above equation onto the reciprocal spherical basis
$\reciprocalSphericalBasisP{lm\sigma}$ yields
\begin{equation}
\label{w projection of flow reflected from two walls}
\langle
   \deltab{a}(i)\reciprocalSphericalBasisP{lm\sigma}(i)
\mid
   \vLout\rangle
+\langle
   \deltab{a}(i)\reciprocalSphericalBasisP{lm\sigma}(i)
\mid
   \vUout\rangle
=\sum_{l'm'\sigma'}
   \GreenWallElement_{ij}(lm\sigma\mid l'm'\sigma')f_j(l'm'\sigma'),
\end{equation}
which is analogous to the single-wall result \refeq{w projection of
flow reflected from single wall}.  Explicit expressions for the matrix
elements $\GreenWallElement_{ij}(lm\sigma\mid l'm'\sigma')$ can thus
be derived by generalizing the analysis presented in Section
\ref{Hyrodynamic interactions with single planar wall}.

To this end, the representation of the velocity fields $\vAin$ in
terms of the Cartesian basis fields
$\CartesianBasisPM{\bk\sigma}(\br_\alpha)$ aligned with the wall
$\alpha$ is obtained by inserting expansions \refeq{multipolar
expansion of scattered field} and \refeq{expansion of v L and U out in
Cartesian basis} into \refeq{incoming and outcoming flows for lower
and upper walls}, and applying the transformation formulas
\refeq{Cartesian displacement theorems -+} and \refeq{factorization
CS}.  Using then the single-wall scattering formula \refeq{definition
of Z wall} to relate the expansion coefficients for the outcoming and
incoming flows we get a pair of coupled linear equations
\begin{subequations}
\label{reflection equations for both wall}
\begin{equation}
\label{reflection equation for lower wall}
\cLout{\bk}=-\ZsingleWall\bcdot[
   \CartesianDisplacement{++}(\bR_{\low\up},\bk)\bcdot\cUout{\bk}
 +\eta^{-1}
  \sum_{l'm'}\CartesianDisplacement{++}(\bR_{\low j},\bk)
\bcdot
   \TransformationCS{+-}(\bk,l'm')
\bcdot
   \inducedForceMultipole(l'm')],
\end{equation}
\begin{equation}
\label{reflection equation for upper wall}
\cUout{\bk}=-\ZsingleWall\bcdot[
   \CartesianDisplacement{--}(\bR_{\up\low},\bk)\bcdot\cLout{\bk}
 +\eta^{-1}
  \sum_{l'm'}\CartesianDisplacement{--}(\bR_{\up j},\bk)
\bcdot
   \TransformationCS{--}(\bk,l'm')
\bcdot
   \inducedForceMultipole(l'm')],
\end{equation}
\end{subequations}
where 
\begin{equation}
\label{relative position vectors}
\bR_{\alpha,\beta}=\bR_\alpha-\bR_\beta\qquad\beta=\low,\up,j.
\end{equation}

In order to to express the solution of the system \refeq{reflection
equations for both wall} in a compact manner we introduce a matrix
notation in the space of six-dimensional vectors of the form
\begin{equation}
\label{six-dimensional vectors c}
\cTwoWalls(\bk)=
\left[
\begin{array}{c}
   \cLout{\bk}\\\\
   \cUout{\bk}
\end{array}
\right].
\end{equation}
Accordingly, we define the $6\times3$ and $3\times6$ transformation
matrices
\begin{subequations}
\label{two wall transformation matrices}
\begin{equation}
\label{two wall C-S transformation matrix}
   \TCS(\bk,lm)=
\left[
\begin{array}{c}
      \TransformationCS{+-}(\bk,lm)
   \\\\
      \TransformationCS{--}(\bk,lm)
\end{array}
\right],
\end{equation}
\begin{equation}
\label{two wall S-C transformation matrix}
   \TSC(lm,\bk)=
\left[
\begin{array}{cc}
      \TransformationSC{+-}(lm,\bk) & \TransformationSC{++}(lm,\bk)
\end{array}
\right],
\end{equation}
\end{subequations}
the $6\times6$ Cartesian displacement matrices
\begin{subequations}
\label{two wall displacement}
\begin{equation}
\label{two wall w-p displacement}
   \Swp{j}(\bk)=
\left[
   \begin{array}{cc}
      \CartesianDisplacement{++}(\bR_{\low j},\bk)&0
\\\\
      0&\CartesianDisplacement{--}(\bR_{\up j},\bk)
   \end{array}
\right],
\end{equation}
\\
\begin{equation}
\label{two wall p-w displacement}
   \Spw{i}(\bk)=
\left[
   \begin{array}{cc}
      \CartesianDisplacement{--}(\bR_{i\low},\bk)&0
\\\\
      0&\CartesianDisplacement{++}(\bR_{i\up},\bk)
   \end{array}
\right],
\end{equation}
\end{subequations}
and the $6\times6$ two-wall flow reflection matrix
\begin{equation}
\label{two wall Z matrix}
   \ZW(\bk)=
\left[
   \begin{array}{cc}
      \ZsingleWall^{-1}&\CartesianDisplacement{++}(\bR_{\low\up},\bk)
\\\\
      \CartesianDisplacement{--}(\bR_{\up\low},\bk)&\ZsingleWall^{-1}
   \end{array}
\right]^{-1}.
\end{equation}
For simplicity, the dependence on the wall and particle positions was
suppressed on the left side of the above expressions.

Due to the symmetries \refeq{symmetry of Cartesian displacement
matrices} and \refeq{CS--SC symmetry} of the $3\times3$ transformation
and displacement matrices, the corresponding symmetry relations
\begin{subequations}
\label{symmetries of two wall matrices}
\begin{equation}
\label{symmetry of two wall transformation matrices}
\TCS(\bk,lm)=[\TSC(lm,\bk)]^\dagger,
\end{equation}
\begin{equation}
\label{symmetry of two wall displacement matrices}
\Swp{i}(\bk)=[\Spw{i}(\bk)]^\dagger,
\end{equation}
\begin{equation}
\label{symmetry of two wall scattering matrices}
\ZW(\bk)=[\ZW(\bk)]^\dagger
\end{equation}
\end{subequations}
are satisfied by the matrices \refeq{two wall transformation
matrices}--\refeq{two wall Z matrix}.  We note that the two-wall
scattering matrix \refeq{two wall Z matrix} involves displacement
components describing translation of the flow field between walls.

Using notation introduced above, the solution of the system
\refeq{reflection equations for both wall} can be written in the
form
\begin{equation}
\label{solution for flow between two walls}
\cTwoWalls(\bk)
   =-\eta^{-1}\sum_{l'm'}
   \ZW(\bk)\bcdot\Swp{j}(\bk)\bcdot\TCS(\bk,l'm')
   \bcdot\inducedForceMultipole_j(l'm').
\end{equation}
In order to get an explicit expression for the matrix elements of the
Green operator $\GreenWall_{ij}$, equations \refeq{expansion of v L
and U out in Cartesian basis}, \refeq{w projection of flow reflected
from two walls}, and \refeq{solution for flow between two walls} are
combined, and the scalar products between the Cartesian and spherical
basis fields are evaluated with the help of relation
\refeq{factorization SC}.  The resulting expression for the elements
of the two-wall Green matrix is
\begin{equation}
\label{two wall Green matrix}
\GreenWall_{ij}(lm\mid l'm')=-\eta^{-1}\int\diff\bk
   \TSC(lm,\bk)\bcdot\Spw{i}(\bk)
      \bcdot\ZW(\bk)
         \bcdot\Swp{j}(\bk)\bcdot\TCS(\bk,l'm').
\end{equation}
The expression is similar in its form (and interpretation) to the
corresponding relation \refeq{expression for single wall G'} for a
single-wall system.  As with the matrix elements \refeq{Fourier
representation of spherical displacement - Green notation} and
\refeq{expression for single wall G'}, the Lorentz symmetry
\refeq{reciprocal relations elements wall Green operator} of the
elements \refeq{two wall Green matrix} is explicit due to the symmetry
relations \refeq{symmetries of two wall matrices} of the component
matrices.

The dependence of the integrand in equation \refeq{two wall Green
matrix} on the polar angles in the Fourier and real spaces is
identical to the corresponding dependence in equation
\refeq{expression for single wall G'}---the angular integration can
thus be performed in a similar manner.  By analogy with equations
\refeq{Fourier representation of single wall Green elements} and
\refeq{expression for g single wall} we get 
\begin{eqnarray}
\label{Fourier representation of two wall Green elements}
   &&\GreenWallElement_{ij}(lm\sigma\mid l'm'\sigma')=
   \eta^{-1}(-1)^{m'-m}
   \e^{\im(m'-m)\phi_{ij}}
\nonumber \\&&
\hphantom{\GreenWallElement_{ij}(lm\sigma}
   \times
      \int_0^{\infty}
      \TwoWallsGreenFourierElement(
         k;lm\sigma\mid l'm'\sigma')
       k^{l+l'+\sigma+\sigma'-2}\BesselJ_{m'-m}(k\rho_{ij})\diff k,
\end{eqnarray}
where
\begin{equation}
\label{expression for g two wall}
\TwoWallsGreenFourier(k;lm\mid l'm')=
   -\tildeTSC(lm)
\bcdot
   \tildeSpw{i}(k)
\bcdot
   \tildeZW(k)
\bcdot
   \tildeSwp{j}(k)
\bcdot
   \tildeTCS(l'm'),
\end{equation}
with the matrices in the product given by
\begin{equation}
\label{two wall transformation matrices scaled}
   \tildeTCS(lm)=[\tildeTSC(lm)]^\dagger=
\left[
\begin{array}{c}
      \tildeTransformationCS{+-}(lm)
   \\\\
      \tildeTransformationCS{--}(lm)
\end{array}
\right],
\end{equation}

\begin{equation}
\label{two wall displacement matrices scaled}
   \tildeSwp{j}(k)=[\tildeSpw{j}(k)]^\dagger=
\left[
   \begin{array}{cc}
      \tildeCartesianDisplacement{++}(kZ_{\low j})&0
\\\\
      0&\tildeCartesianDisplacement{--}(kZ_{\up j})
   \end{array}
\right],
\end{equation}
and 
\begin{equation}
\label{two wall Z matrix scaled}
   \tildeZW(k)=
\left[
   \begin{array}{cc}
      \ZsingleWall^{-1}&\tildeCartesianDisplacement{++}(-kH)
\\\\
      \tildeCartesianDisplacement{--}(kH)&\ZsingleWall^{-1}
   \end{array}
\right]^{-1},
\end{equation}
where
\begin{equation}
\label{distance between walls}
H=Z_\up-Z_\low
\end{equation}
is the wall separation.  The above relations, together with equation
\refeq{expression for tilde S}, imply that
\mbox{$\TwoWallsGreenFourier(k;lm\mid l'm')$} depends on the
$z$-coordinates of the walls and the points $i$ and $j$, but is
independent of the lateral coordinates, consistently with the
translational invariance of the system.  Since the two-wall scattering
matrix \refeq{two wall Z matrix scaled} is more complex than its
one-particle counterpart, the integration in equation \refeq{Fourier
representation of two wall Green elements} (unlike \refeq{Fourier
representation of single wall Green elements}) cannot be analytically
performed.  However, numerical integration is straightforward, except
when the lateral distance between points $i$ and $j$ is large, in
which case the oscillatory character of the integrand becomes
important.

\section{Stokesian-dynamics algorithm for suspension between two walls}
\label{Numerical implementation}

The theoretical results derived in the previous sections enable
development of efficient numerical algorithms for evaluation of
many-body hydrodynamic interactions in suspensions of spherical
particles confined between two planar walls.  To our knowledge, such
algorithms have not been available so far.  In what follows, we
describe a many-particle Stokesian-dynamics algorithm based on our
theory.

In Section \ref{Friction matrix} we summarize expressions that relate the
matrix $\GrandMobility$ in the force-multipole equation \refeq{induced
force equations in matrix notation} to the resistance matrix in a
suspension of many spheres.  Our transformation formulas relating the
spherical and Cartesian basis fields are employed in Section
\ref{Integrands}, where a simple numerical-integration procedure for
evaluating the elements of the matrix $\GrandMobility$ is described.
The lubrication-subtraction techniques
\cite{Durlofsky-Brady-Bossis:1987,%
Cichocki-Felderhof-Hinsen-Wajnryb-Blawzdziewicz:1994,%
Cichocki-Jones-Kutteh-Wajnryb:2000%
} 
used for improving convergence with the order of the force multipoles
included in the calculation are outlined in Section \ref{Convergence on l}.
Examples of numerical results for the friction matrix of a single
particle, a pair of particles, and in many-particle systems are
presented in Section \ref{Results}.

\subsection{Resistance  matrix} 
\label{Friction matrix}

We focus on a system of $N$ spheres undergoing translational and
rotational rigid-body motion \refeq{rigid-body velocity of drop i}
with no external flow. The particle dynamics in the system is
characterized by the resistance matrix
\begin{equation}
\label{resistance matrix}
\resistanceMatrix_{ij}=
\left[
  \begin{array}{cc}
    \resistanceMatrixTT_{ij}&\resistanceMatrixTR_{ij}\\
    \resistanceMatrixRT_{ij}&\resistanceMatrixRR_{ij}
  \end{array}
\right],
\qquad i,j=1,\ldots,N,
\end{equation}
defined by the linear relation
\begin{equation}
\label{resistance relation}
\left[
  \begin{array}{c}
\totForce_i\\
\totTorque_i
  \end{array}
\right]
=
\sum_{j=1}^N
\left[
  \begin{array}{cc}
    \resistanceMatrixTT_{ij}&\resistanceMatrixTR_{ij}\\
    \resistanceMatrixRT_{ij}&\resistanceMatrixRR_{ij}
  \end{array}
\right]
\bcdot
\left[
  \begin{array}{c}
     \bU_j\\
     \bOmega_j
  \end{array}
\right]
\end{equation}
between the translational and rotational particle velocities $\bU_j$
and $\bOmega_j$ and the forces and torques $\totForce_i$ and
$\totTorque_i$ applied to the particles.  The dot in equation
\refeq{resistance relation} denotes the matrix multiplication and
contraction of the Cartesian tensorial components of the resistance
matrix.  A detailed discussion of a more general resistance problem,
which involves an external linear flow and the stresslet induced on
the particles, is presented in
Ref.~\cite{Cichocki-Jones-Kutteh-Wajnryb:2000}.

The resistance relation \refeq{resistance relation} can be linked to
the induced-force equation \refeq{induced force equations for matrix
elements} by expressing the applied forces and torques $\totForce_i$
and $\totTorque_i$ in terms of the induced-force distributions
\refeq{induced forces},
\begin{equation}
\label{force and torque}
   \totForce_i=\int\bF_i(\br)\diff\br,
\qquad
   \totTorque_i=\int\br_i\btimes\bF_i(\br)\diff\br.
\end{equation}
Representing the above quantities in terms of the induced-force
multipoles \refeq{induced force multipoles} yields
\begin{equation}
\label{force and torque projections}
\totForce_i=\sum_{lm\sigma}
   \frictionProjectionVector(\transl\mid lm\sigma)f(lm\sigma),
\qquad
\totTorque_i=\sum_{lm\sigma}
   \frictionProjectionVector(\rot\mid lm\sigma)f(lm\sigma),
\end{equation}
where
\begin{subequations}
\label{friction projection matrix}
\begin{equation}
\label{friction projection matrix - translational}
\frictionProjectionVector(\transl\mid lm\sigma)
   =\delta_{l1}\delta_{\sigma0}\tilde\frictionProjectionVector^\transl(m),
\end{equation}
\begin{equation}
\label{friction projection matrix - rotational}
\frictionProjectionVector(\rot\mid lm\sigma)
   =\delta_{l1}\delta_{\sigma1}\tilde\frictionProjectionVector^\rot(m).
\end{equation}
\end{subequations}
Explicit expressions for the vectors
$\tilde\frictionProjectionVector^\transl(m)$ and
$\tilde\frictionProjectionVector^\rot(m)$ are listed in Appendix
\ref{Transformation vectors X}.
The coefficients $c_i$ in the corresponding expansion \refeq{expansion
of external flow} for the flow field \refeq{rigid-body velocity of
drop i} can be represented in the form 
\begin{equation}
\label{expansion of rigid-body velocity field}
c_i(lm\sigma)
  =\frictionProjectionVector(lm\sigma\mid\transl)\bcdot\bU_i
  +\frictionProjectionVector(lm\sigma\mid\rot)\bcdot\bOmega_i,
\end{equation}
where $\externalVelocity(\br)=0$ is assumed.  As shown below,
the projection matrices $\frictionProjectionVector$ in equations
\refeq{force and torque projections} and \refeq{expansion of
rigid-body velocity field} obey the identity
\begin{equation}
\label{symmetry of friction projection matrices}
\frictionProjectionVector(lm\sigma\mid A)
   =\frictionProjectionVector^*(A\mid lm\sigma),
\qquad
   A=\transl,\rot.
\end{equation}

In order to determine the resistance matrix $\resistanceMatrix_{ij}$
from the above expressions, the force-multipole equation
\refeq{induced force equations in matrix notation} is solved, which
yields the linear relation
\begin{equation}
\label{solution of equation for force multipoles}
\inducedForceMultipole_i(lm)
   =\sum_{j=1}^N\sum_{l'm'}
      \GrandFriction_{ij}(lm\mid l'm')
   \bcdot
      \externalVelocityCoefficient_j(l'm'),
\end{equation} 
where $\GrandFriction=\GrandMobility^{-1}$ is the generalized friction
matrix.  By inserting into equation \refeq{solution of equation for
force multipoles} the projection formulas \refeq{force and torque
projections} and \refeq{expansion of rigid-body velocity field} we get
\begin{equation}
\label{Elements of physical friction matrix}
\resistanceMatrix^{AB}_{ij}
   =\sum_{lm\sigma}\sum_{l'm'\sigma'}
      \frictionProjectionVector(A\mid lm\sigma)
         \GrandFrictionElement_{ij}(lm\sigma\mid l'm'\sigma')
            \frictionProjectionVector(l'm'\sigma'\mid B),
\end{equation}
where $A,B=\transl,\rot$.  

With our normalization of the spherical basis fields \refeq{spherical
basis v -} and \refeq{spherical basis v +} (as defined in Section
\ref{spherical basis}) the symmetry relation \refeq{symmetry of
friction projection matrices} is a direct consequence of the Lorentz
symmetry of the generalized friction matrix
\begin{equation}
\label{Lorentz symmetry of generalized friction matrix}
\GrandFrictionElement_{ij}(lm\sigma\mid l'm'\sigma')
   =\GrandFrictionElement_{ji}(l'm'\sigma'\mid lm\sigma),
\end{equation}
which follows from equations \refeq{reciprocal relations elements of
Oseen operator}--\refeq{reciprocal relations elements of one-particle
scattering operator}.  Relation \refeq{symmetry of friction projection
matrices} is obtained by inserting equation \refeq{Lorentz symmetry of
generalized friction matrix} into \refeq{Elements of physical friction
matrix} and using the Lorentz symmetry of the resistance matrix
\cite{Kim-Karrila:1991}
\begin{equation}
\label{Lorentz symmetry of resistance matrix}
\resistanceMatrix^{AB}_{ij}
   =\left[\resistanceMatrix^{B A}_{ji}\right]^\dagger,
\end{equation}
where the dagger denotes the transposition of a tensor.

\subsection{Evaluation of matrix $\GrandMobility$}
\label{Integrands}

The evaluation of the resistance matrix $\resistanceMatrix^{AB}_{ij}$
from expression \refeq{Elements of physical friction matrix} requires
solving the set of linear algebraic equations for the induced-force
multipoles \refeq{induced force equations in matrix notation} to
obtain the generalized friction coefficients
$\GrandFrictionElement_{ij}(lm\sigma\mid l'm'\sigma')$.  The matrix
$\GrandMobility$ in the equation \refeq{induced force equations in
matrix notation} is given as the sum of three terms \refeq{Grand
Mobility matrix}.  The first two terms, i.e., the single-particle
scattering matrix $\Zsingle_i$ and the matrix $\GreenFree_{ij}$
representing the flow in infinite space, are known explicitly
\cite{Felderhof-Jones:1989,Cichocki-Felderhof-Schmitz:1988}.  The
remaining term---the two-wall contribution
$\GreenWall_{ij}$---is evaluated numerically, using relations
\refeq{Fourier representation of two wall Green elements}--\refeq{two
wall Z matrix scaled} along with our expressions for the Cartesian
displacement matrices \refeq{expression for tilde S}, the
transformation matrices \refeq{form of transformation SC and CS}, and
the single-wall scattering matrix \refeq{expression for Z wall}.

\begin{figure}

\begin{center}
\includegraphics{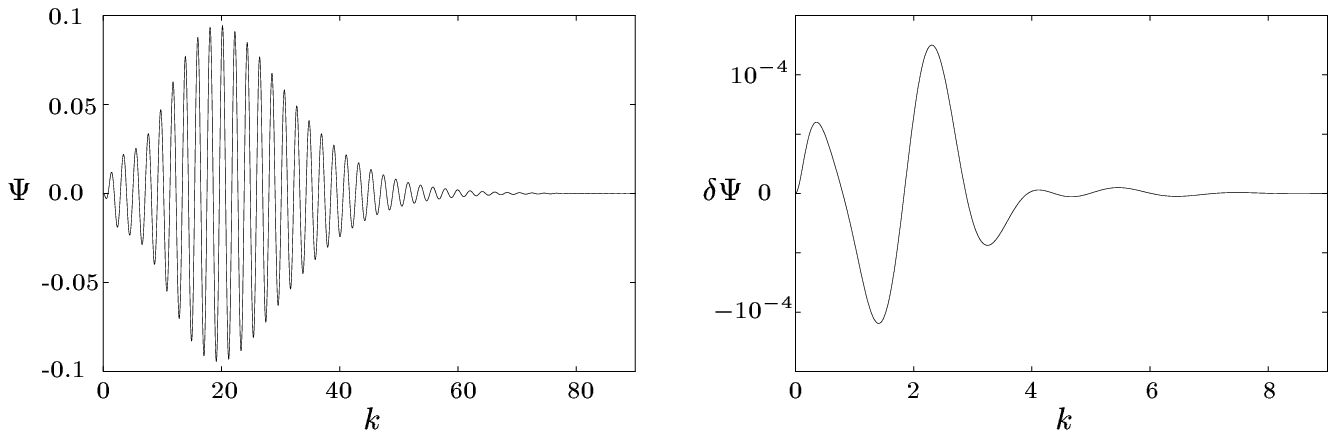}
\end{center}

\input{figtex/Integrand-cap}
\end{figure}


As already mentioned at the end of Section \ref{Suspension between two
planar walls--matrix elements}, numerical evaluation of the integral
\refeq{Fourier representation of two wall Green elements} is
straightforward for sufficiently small lateral separations between
particles $i$ and $j$.  For large interparticle separations
$\rho_{ij}$, however, the integration is more difficult because of the
oscillatory behavior of the integrands
\begin{equation}
\label{two wall integrand}
\integrandTwoWall(k)=
      \TwoWallsGreenFourierElement(
         k;lm\sigma\mid l'm'\sigma')
       k^{l+l'+\sigma+\sigma'-2}\BesselJ_{m'-m}(k\rho_{ij}),
\end{equation}
due to the presence of the factor $\BesselJ(k\rho_{ij})$.  This
behavior is illustrated in the left panel of figure \ref{oscillatory
integrands} for a configuration in which both points $i$ and $j$ are
close to one of the walls.

To avoid numerical integration of
a highly oscillatory function, the integrand \refeq{two wall
integrand} is decomposed
\begin{equation}
\label{decomposition of integrand}
\Psi(k)=\Psi_\low(k)+\Psi_\up(k)+\delta\Psi(k)
\end{equation}
into the superposition of the single-wall contributions $\Psi_\low$ and
$\Psi_\up$, and the wall-interaction part $\delta\Psi$.  According to
equations \refeq{Fourier representation of single wall Green elements}
and \refeq{expression for g single wall}, the single-wall integrands
are
\begin{equation}
\label{single-wall integrand}
\Psi_\alpha(k)=
      \SingleWallGreenFourierElementPM(
         kZ_{i\alpha},kZ_{\alpha j};lm\sigma\mid l'm'\sigma')
         k^{l+l'+\sigma+\sigma'-2}\BesselJ_{m'-m}(k\rho_{ij}),
\end{equation}
where $\alpha=\low,\up$.  Relation \refeq{exponential factor for
single wall} implies that for large values of $k$ the amplitude of
these integrands decays as
\begin{equation}
\label{asymptotic behavior of integrand for wall alpha}
\Psi_\alpha(k)\sim\e^{-k\imageOffset_{ij}^{(\alpha)}},
\end{equation}
 where $\imageOffset_{ij}^{(\alpha)}$ is the vertical offset
\refeq{offset - one wall} between the point $i$ and the reflection of
point $j$ in the wall $\alpha$.  Therefore, the decay is slow if both
points $i$ and $j$ are close to the wall, consistent with the results
in the left panel of figure \ref{oscillatory integrands}.  

In contrast, the decay of the wall-interaction part $\delta\Psi(k)$ of
integrand \refeq{decomposition of integrand} is determined by the wall
separation $H$.  As shown in Appendix \ref{Large k behavior of
integrands delta Psi}, the large-$k$ asymptotic behavior of this
function is
\begin{equation}
\label{asymptotic behavior of wall-interaction integrand}
\delta\Psi(k)\sim\e^{-k{\tilde\imageOffset}_{ij}},
\end{equation}
where
\begin{equation}
\label{distance to second image}
{\tilde\imageOffset}_{ij}=2H-|Z_{ij}|>H.
\end{equation}
The lengthscale ${\tilde\imageOffset}_{ij}$ equals the vertical offset
$|Z_i-Z_j''|$ between the target point $i$ and the closer of the two
second-order images of the source point $j$.  The images are at the
positions
\begin{equation}
\label{position of second-order image}
\bR_j''=\bR_j\pm2H\ez,
\end{equation}
where the plus sign corresponds to reflecting the source point first
in the lower wall and then in the upper wall; the minus sign
corresponds to the opposite order of reflections.

A typical form of the wall-interaction contribution $\delta\Psi(k)$ is
presented in the right panel of figure \ref{oscillatory integrands}.
Unlike the results for $\Psi(k)$, the integrand $\delta\Psi(k)$ in
this example is negligible already after several oscillations.  Thus,
the function $\delta\Psi(k)$ is easy to integrate numerically.

In our algorithm, the contribution
$\delta\GreenWallElement_{ij}(lm\sigma\mid l'm'\sigma')$ to the matrix
elements \refeq{Fourier representation of two wall Green elements},
associated with the component $\delta\Psi(k)$ of the integrand, is
evaluated by numerical integration using the Simpson rule.  The slowly
convergent one-wall contributions \refeq{single-wall integrand} are
calculated analytically, using the explicit image-representation
expressions \cite{Cichocki-Jones:1998} (cf., the discussion in Section
\ref{Matrix elements of Green operator - single wall}).

Our numerical tests indicate that this procedure yields accurate
results for $\rho_{ij}/H\lesssim20$.  The procedure can be further
improved, either by subtracting several terms associated with
higher-order wall reflections of the source multipole
\cite{Bhattacharya-Blawzdziewicz:2002}, or by deriving asymptotic
formulas for the integrals \refeq{Fourier representation of two wall
Green elements}.

\subsection{Convergence with multipolar order}
\label{Convergence on l}

\begin{figure}

\begin{center}
\includegraphics{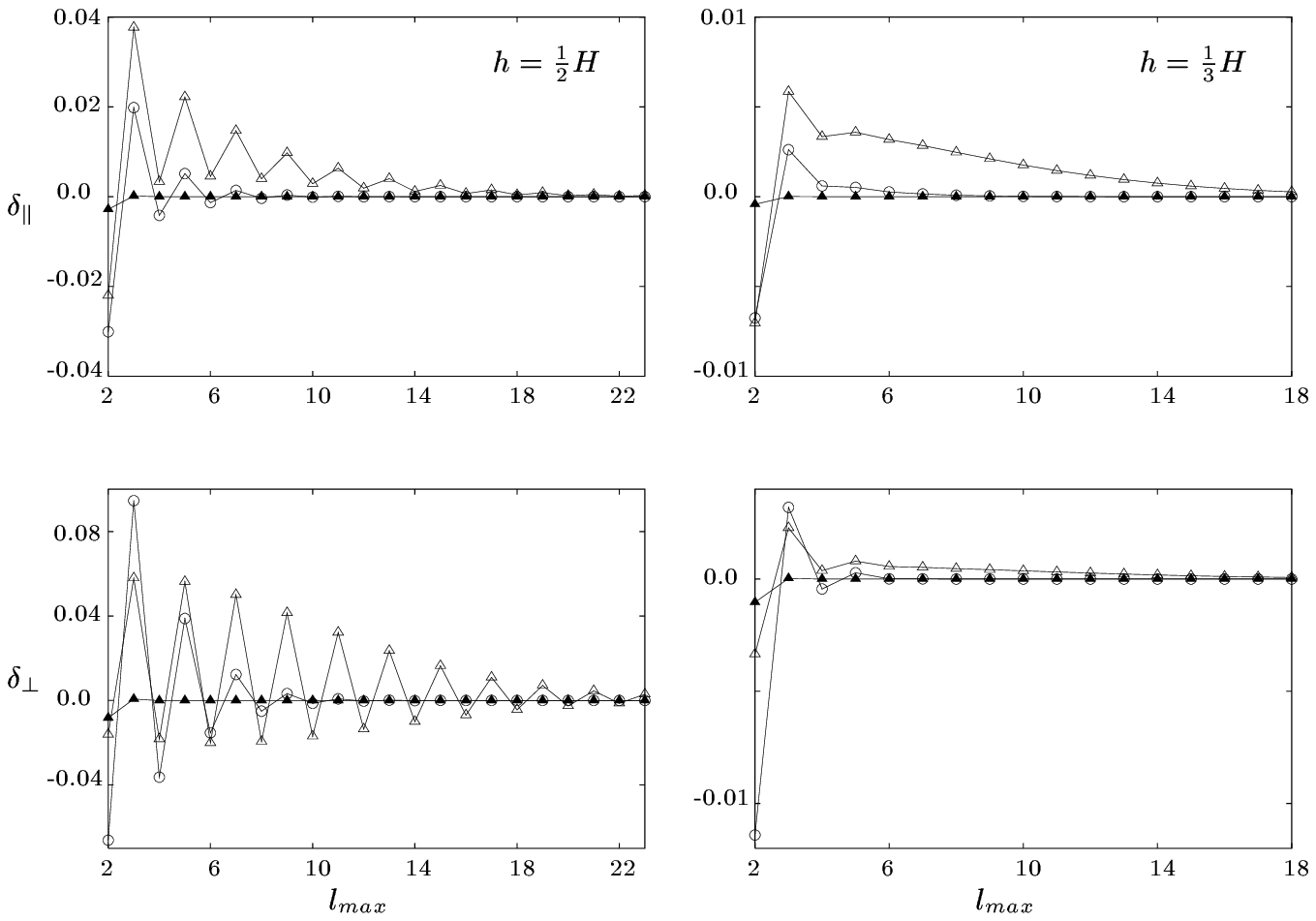}
\end{center}

\input{figtex/Convergence-l-cap}
\end{figure}

Our numerical procedure for evaluating the friction matrix involves
truncation of the linear system \refeq{induced force equations in
matrix notation} by neglecting the induced force multipoles
$\inducedForceMultipole_i(lm)$ of the order $l>\lmax$.  This
multipolar approximation converges very slowly with the truncation
order $\lmax$ if any two particles are close together or a particle is
close to a wall.  Such a behavior stems from a rapid variation of the
flow field in the near-contact lubrication regions.  The mechanism is
well known and has been observed for particles in infinite space and
in the presence of a single wall.
To overcome this difficulty we employ a standard method, originally
introduced by \cite{Durlofsky-Brady-Bossis:1987}, according to which
the lubrication forces are included in the friction matrix using a
superposition approximation.  We follow closely the implementation of
the method described by \cite{Cichocki-Jones-Kutteh-Wajnryb:2000} in
their study of a single wall problem.  Accordingly, the resistance
matrix \refeq{resistance matrix} is represented in the form
\begin{equation}
\label{lubrication superposition for resistance matrix}
\resistanceMatrix_{ij}=
    \resistanceMatrixSupPP_{ij}
      +\resistanceMatrixSupWP_{ij}
      +\Delta\resistanceMatrix_{ij},
\end{equation}
where 
\begin{equation}
\label{two-particle superposition}
\resistanceMatrixSupPP_{ij}
   =\delta_{ij}
      \sum_{\stackrel{\scriptsize\mbox{$k=1$}}{k\not=i}}^N
         \resistanceMatrixTPNoWall{ii}(ik)
   +(1-\delta_{ij})\resistanceMatrixTPNoWall_{ij}(ij)
\end{equation}
and
\begin{equation}
\label{particle-wall superposition}
\resistanceMatrixSupWP_{ij}
   =\delta_{ij}[\resistanceMatrix^{\low}_{i}(i)
   +\resistanceMatrix^{\up}_{i}(i)].
\end{equation}
Here $\resistanceMatrixTPNoWall_{mm}(mn)$ is the self- and
$\resistanceMatrixTPNoWall_{mn}(mn)$ the mutual-resistance matrix for
an isolated pair of particles $m$ and $n$ in the unbounded space, and
$\resistanceMatrix^{\alpha}_{m}(m)$ is the single-particle resistance
matrix for a sphere in the subspace bounded by the wall
$\alpha=\low,\up$.  The superposition contributions
\refeq{two-particle superposition} and \refeq{particle-wall
superposition} in equation \refeq{lubrication superposition for
resistance matrix} can be calculated with high accuracy, using methods
discussed below.  The convergence with the truncation order $\lmax$ of
the multipolar approximation
\begin{equation}
\label{multipolar approximation for zeta}
\resistanceMatrix_{ij}\approx
    \resistanceMatrixSupPP_{ij}
      +\resistanceMatrixSupWP_{ij}
      +[\Delta\resistanceMatrix_{ij}]_\lmax
\end{equation}
is much faster than the convergence of the multipolar approximation
$[\resistanceMatrix_{ij}]_\lmax$ itself, where $[B]_\lmax$ denotes the
quantity $B$ evaluated using the multipolar expansion truncated at
$l=\lmax$. Therefore the evaluation procedure based on equation
\refeq{multipolar approximation for zeta} yields accurate results for
the friction matrix at a substantially reduced numerical cost.

In our implementation, the two-particle components
$\resistanceMatrixTPNoWall_{mm}(mn)$ and
$\resistanceMatrixTPNoWall_{mn}(mn)$ of the friction matrix in the
superposition formula \refeq{two-particle superposition} are evaluated
using a combination of the lubrication resistance expressions
\cite{Kim-Karrila:1991} and the series expansion in inverse powers
of interparticle separation
\cite{Cichocki-Felderhof-Schmitz:1988}.  Similarly, the
one-particle friction matrix $\resistanceMatrix^\alpha_{m}(m)$ in the
superposition formula \refeq{particle-wall superposition} is evaluated
using a combination of the lubrication resistance expression and the
power series in the inverse distance of the particle from the wall
\cite{Cichocki-Jones:1998}.

Our numerical results indicate that for large and moderate wall
separations $H$ (compared to the particle diameter) the multipolar
approximation \refeq{multipolar approximation for zeta} converges
rapidly with the truncation order $\lmax$. For configurations with
$H\approx 2a$ the convergence is less satisfactory, particularly for
the transverse components of the friction matrix.  This behavior is
illustrated in figure \ref{figure of convergence with l}, where the
relative error for the lateral and vertical components
\begin{equation}
\label{single particle lateral and vertical friction matrix}
\onePartResistanceLat
   =
      \resistanceMatrixElement^{\transl\transl\,xx}_{11}
   =
      \resistanceMatrixElement^{\transl\transl\,yy}_{11},
\qquad
\onePartResistanceTrans
   =\resistanceMatrixElement^{\transl\transl\,zz}_{11}
\end{equation}
of the one-particle translational friction matrix is shown for
different particle configurations and distances between the walls.
The results are given for the center and an off-center position of the
particle in the space between the walls,
\refstepcounter{equation}
$$
\label{center and off-center positions}
h=\half H,\qquad h=\third H,
\eqno{(\theequation{\mathit{a},\mathit{b}})}
$$ 
where $h$ is the distance of the particle from the lower wall.  In the
case of the center particle position (\ref{center and off-center
positions}\textit{a}), the multipole-truncation error exhibits
decaying oscillations.  For small values of the gap
\begin{equation}
\label{particle-wall gap normalized by a}
\PWgap=h/a-1
\end{equation}
between the particle surface and the closer wall a typical error is in
the range of several percent.  The results corresponding to
truncations at even orders of $\lmax$ are more accurate than the
results corresponding to truncations at odd orders.  The
multipolar-truncation error for the off-center particle position
(\ref{center and off-center positions}\textit{b}) is much smaller than
the error for the center configuration with the same values of the
particle--wall distance $h$.

A similar dependence of the truncation error on $\lmax$ was observed
for many-particle friction matrix: the error is small, except when the
wall separation is comparable to the particle diameter.  This behavior
suggests that the relatively large error for such tight configurations
results from an interaction between the lubrication layers---this
effect is not accounted for in the superposition terms in equation
\refeq{multipolar approximation for zeta}. The problem, however,
requires further investigations in order to develop better
approximation methods.

\begin{figure}
\begin{center}
\scalebox{0.94}{\includegraphics{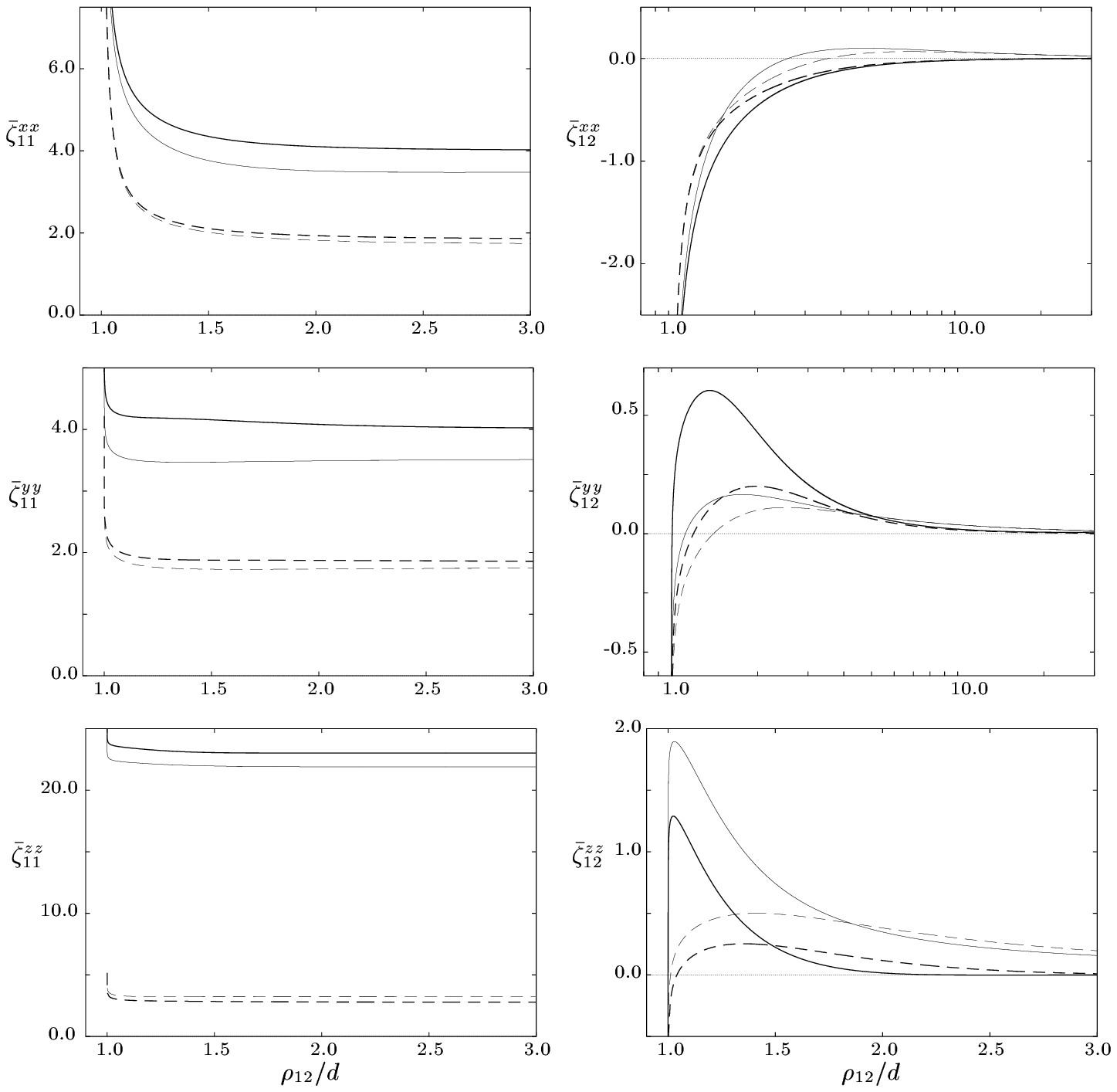}}
\end{center}

\input{figtex/twoPart-New-cap}
\end{figure}
\begin{figure}
\begin{center}

\scalebox{0.94}{\includegraphics{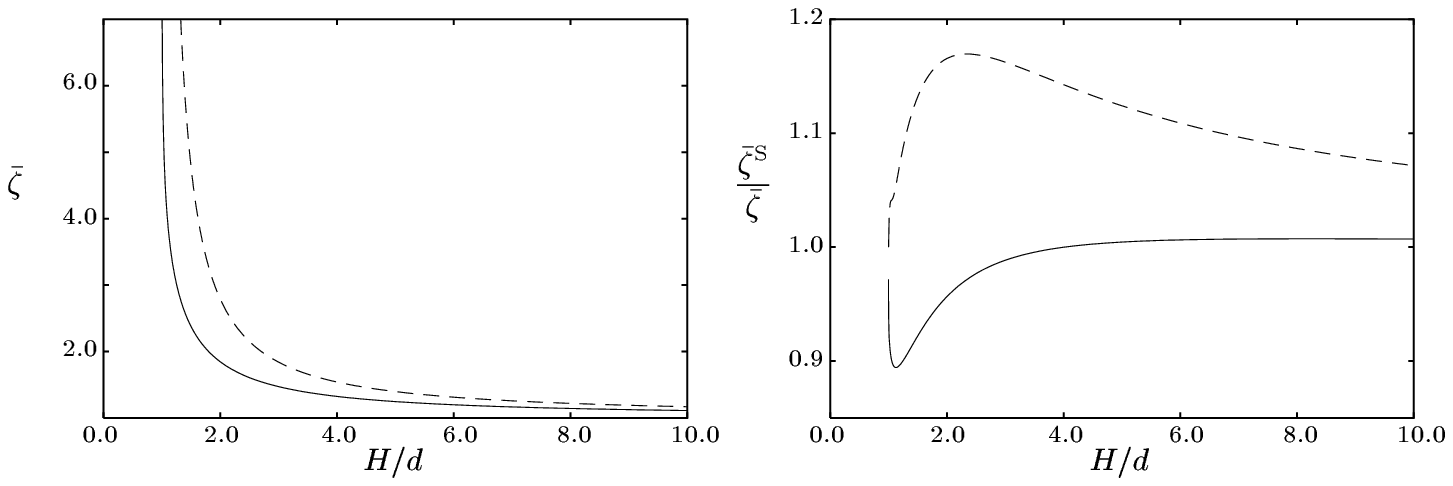}}
\end{center}

\input{figtex/singlePart-New-cap}
\end{figure}
\begin{figure}

\begin{center}
\scalebox{0.94}{\includegraphics{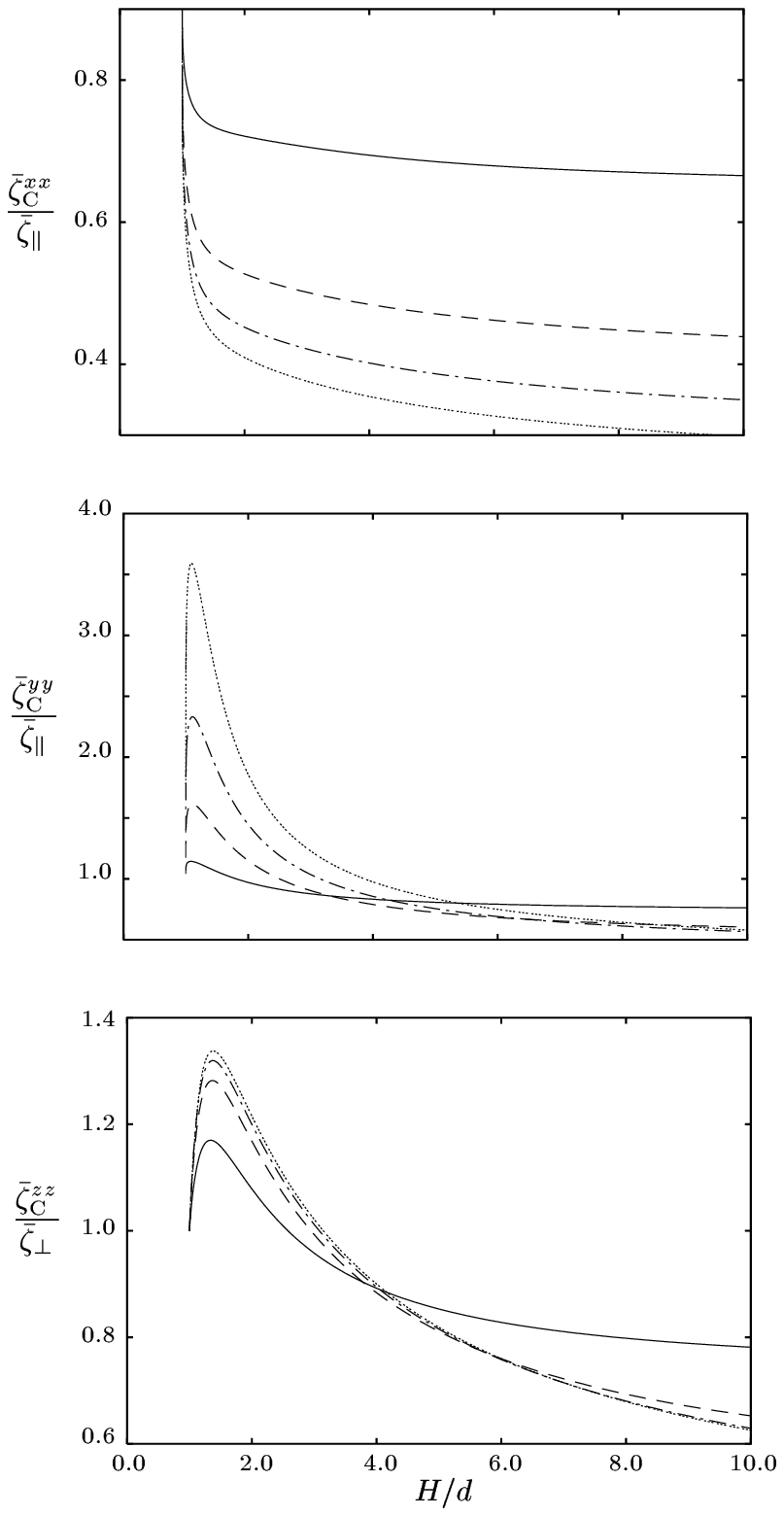}}
\end{center}

\input{figtex/polymer-gap-New-cap}
\end{figure}
\begin{figure}

\begin{center}
\scalebox{0.94}{\includegraphics{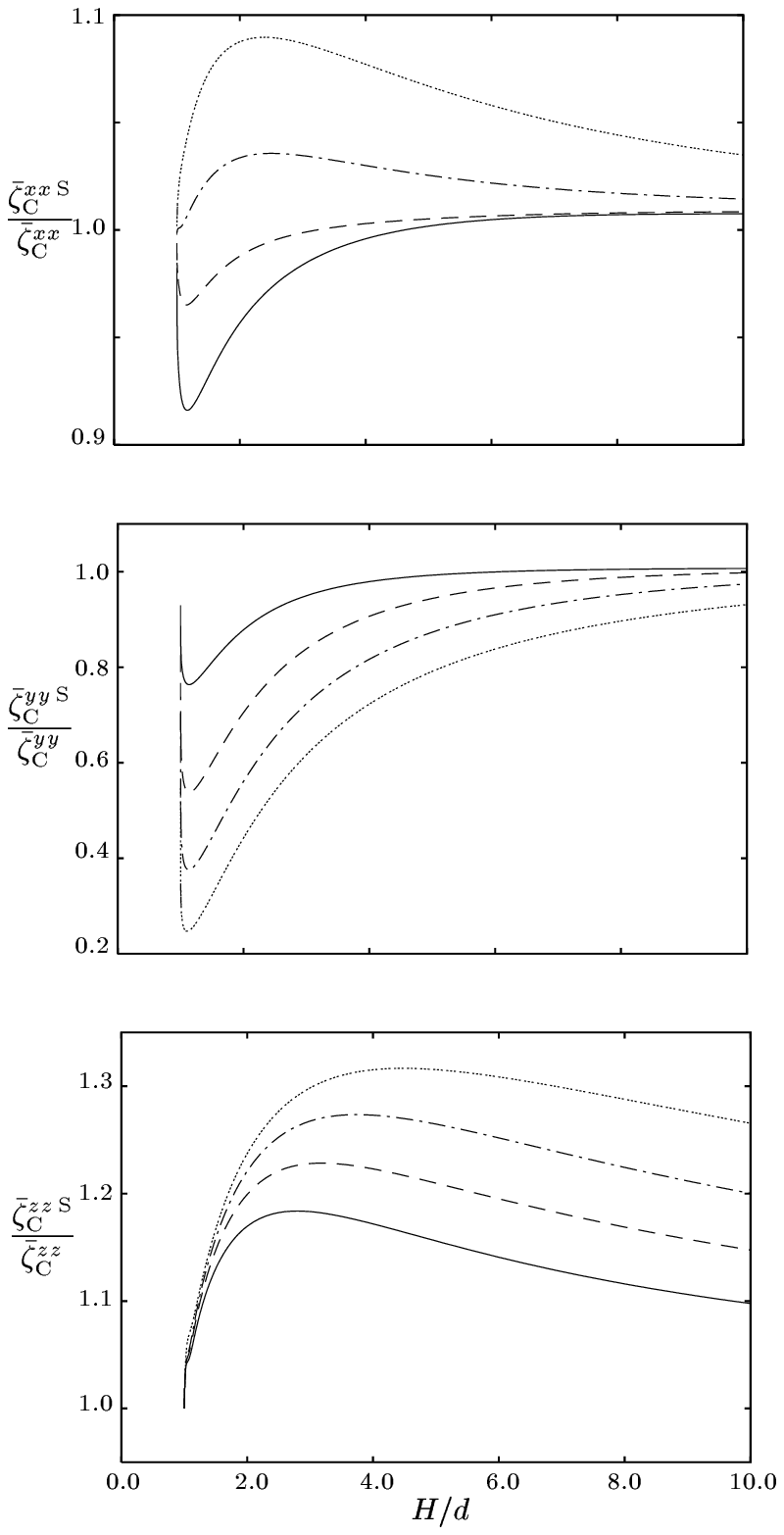}}
\end{center}

\input{figtex/polymer-SuperAccuracy-New-cap}
\end{figure}

\section{Numerical results}
\label{Results}

In this section we give some examples of numerical results for the
hydrodynamic friction matrix in systems of spherical particles
confined between two parallel planar walls.  The calculations for a
single particle and for particle pairs, depicted in Figs.~\ref{new two
particles} and \ref{new single particle}, were performed using the
multipolar approximation \refeq{multipolar approximation for zeta}
with the truncation at the order $\lmax=12$.  The multi-particle
calculations, depicted in Figs.~\ref{new polymer gap} and \ref{Super
Accuracy}, were obtained using $\lmax=8$. These truncations are
sufficient to obtain results with the accuracy better than the
resolution of the plots.  A more extensive set of numerical results is
presented in a separate publication
\cite{Bhattacharya-Blawzdziewicz-Wajnryb:2005}.

\subsection{Two-particle friction matrix}
\label{Two-particle friction matrix}

Figure \ref{new two particles} illustrates the behavior of the
translational components of the two-particle resistance matrix,
normalized by the Stokes friction coefficient
$\StokesResistance=6\upi\eta a$,
\begin{equation}
\label{normalized two-particle resistance matrix}
\twoParticleResistanceNorm_{ij}^{\alpha\beta}
   =\resistanceMatrixElement_{ij}^{\transl\transl\,\alpha\beta}
    /\StokesResistance,\qquad i,j=1,2,
\end{equation}
where $\alpha,\beta=x,y,z$.  The
particle pair is in the center plane of the space between the walls
\begin{equation}
\label{center particle configuration}
h_1=h_2=\half H,
\end{equation}
where $h_i$ is the distance of particle $i$ from the lower wall.  
The relative particle displacement is along the $x$ direction
\begin{equation}
\label{relative position along axis x}
\PPvector=\PPdistance\ex,
\end{equation}
and the results are plotted versus the interparticle distance
$\PPdistance$.  Only the diagonal Cartesian components
$\twoParticleResistanceNorm^{\alpha\alpha}_{11}$ and
$\twoParticleResistanceNorm^{\alpha\alpha}_{12}$ of the self- and
mutual-resistance matrices are shown, because
$\twoParticleResistanceNorm^{\alpha\beta}_{ij}=0$ for
$\alpha\not=\beta$, due to symmetry.

\subsubsection{Near-contact and intermediate behavior}
\label{Behavior at small and intermediate distances}

According to the results shown in the left panels of Fig.~\ref{new two
particles}, the self-components of the two-particle resistance matrix
are only weakly affected by the presence of the second particle,
except for sufficiently small gaps between the particle surfaces
\begin{equation}
\label{interparticle gap}
\PPgap=\PPdistance/d-1,
\end{equation}
where $d=2a$ is the particle diameter.  The effect of the
interparticle interactions is most pronounced for the longitudinal
component $\twoParticleResistanceNorm^{xx}_{11}$, because of the
strong $O(\PPgap^{-1})$ lubrication resistance for particles in
relative motion along the line connecting their centers.  For the
motion in the transverse directions $y$ and $z$, a significant
interparticle-interaction effect is seen only for very small
interparticle gaps, because the transverse interparticle lubrication
resistance has a much weaker logarithmic singularity $O(\log\PPgap$)
than the longitudinal one.

The results in the right panels of Fig.~\ref{new two particles}
indicate that for small interparticle distances all three components
$\twoParticleResistanceNorm^{\alpha\alpha}_{12}$, $\alpha=x,y,z$, of
the mutual friction matrix are negative.  The negative sign indicates
that the hydrodynamic force $\totForce_1^{\rm H}=-\totForce_1$
produced on particle $(1)$ by the motion of particle $(2)$ points in
the same direction as the particle velocity.  When the distance
between particles is increased, the transverse components
$\twoParticleResistanceNorm^{yy}_{12}$ and
$\twoParticleResistanceNorm^{zz}_{12}$ change sign, which results from
the backflow associated with the flow field scattered from the walls.
In contrast, the longitudinal component
$\twoParticleResistanceNorm^{xx}_{12}$ remains negative.  We note that
the backflow effect does not occur in the unbounded space.

\subsubsection{Far-field behavior}
\label{Far-field behavior}

At large interparticle separations $\PPdistance/d\gg1$, the mutual
components $\twoParticleResistanceNorm^{\alpha\alpha}_{12}$ of the
two-particle resistance matrix vanish, and the self-components
$\twoParticleResistanceNorm^{\alpha\alpha}_{11}$ tend to the
corresponding one-particle values 
\begin{equation}
\label{normalized lateral and vertical single-particle friction}
\baronePartResistanceLat=\onePartResistanceLat/\StokesResistance,
\qquad
\baronePartResistanceTrans=\onePartResistanceTrans/\StokesResistance
\end{equation}
($\baronePartResistanceLat$ for
$\twoParticleResistanceNorm^{xx}_{11}$ and
$\twoParticleResistanceNorm^{yy}_{11}$, and
$\baronePartResistanceTrans$ for
$\twoParticleResistanceNorm^{zz}_{11}$).  The lateral and transverse
one-particle friction coefficients $\baronePartResistanceLat$ and
$\baronePartResistanceTrans$ are shown in the left panel of
Fig.~\ref{new single particle} versus the wall separation $H$ for the
center particle position $h=\half H$.  Our present one-particle
results agree with those of Jones \cite{Jones:2004} and with our
earlier solution obtained using the image-singularity technique
\cite{Bhattacharya-Blawzdziewicz:2002}.

The asymptotic approach of the two-particle friction matrix
$\twoParticleResistanceNorm_{ij}^{\alpha\alpha}$ to the limiting
values at large $\PPdistance$ can be determined from the far-field
behavior of the flow field $\bv^{\rm as}$ produced by a particle moving
in the horizontal direction $\be_\alpha$ ($\alpha=x,y$).  By expanding
the flow $\bv^{\rm as}$ in the small parameter $H/\rho$, where $\rho\gg
H$ is the horizontal distance from the moving particle, we find that
\begin{equation}
\label{far-field flow produced by a moving particle}
\bv^{\rm as}=\half\eta^{-1} z(H-z)\bnabla p^{\rm as}.
\end{equation}
Here $z=0$ is the position of the lower wall, and 
\begin{equation}
\label{far field pressure}
p^{\rm as}\sim\frac{\brho\bcdot\be_\alpha}{\rho^2},\qquad\alpha=x,y,
\end{equation}
is the pressure field that depends only on the lateral position
$\brho=x\ex+y\ey$.  The above equation indicates that the far-field
velocity \refeq{far-field flow produced by a moving particle} decays
as
\begin{equation}
\label{decay of far field flow}
\bv^{\rm as}\sim \rho^{-2}
\end{equation}
for $\rho\to\infty$.  One can also show that the flow field $\bv^{\rm
as}$ produced by a particle moving in the $z$ direction decays
exponentially.  The above results are consistent with the asymptotic
expression for the flow field produced in the space between the walls
by a Stokeslet pointing in the horizontal direction
\cite{Liron-Mochon:1976}.  (A general analysis of the far-field flow
will be presented elsewhere.)

The result \refeq{decay of far field flow} implies that the asymptotic
far-field behavior of the lateral components of the friction matrix is
\begin{equation}
\label{large rho behavior of lateral self friction matrix}
\twoParticleResistanceNorm_{11}^{\alpha\alpha}
   =\baronePartResistanceLat+O(\PPdistance^{-4}),
\end{equation}
\begin{equation}
\label{large rho behavior of lateral mutual friction matrix}
\twoParticleResistanceNorm_{12}^{\alpha\alpha}=O(\PPdistance^{-2}),
\end{equation}
where $\alpha=x,y$ (the $O(\PPdistance^{-4})$ contribution in
Eq.~\refeq{large rho behavior of lateral self friction matrix}
corresponds to the flow field \refeq{decay of far field flow}
scattered back to the original particle).  In contrast to the results
\refeq{large rho behavior of lateral self friction matrix} and
\refeq{large rho behavior of lateral mutual friction matrix}, the 
limits
$\twoParticleResistanceNorm_{11}^{zz}=\baronePartResistanceTrans$ and
$\twoParticleResistanceNorm_{12}^{zz}=0$ for the vertical components
of the friction matrix are approached exponentially on the lengthscale
$H$.  The numerical results shown in Fig.~\ref{new two particles}
agree with the above analysis.  In particular, the signs of the
longitudinal and transverse friction coefficients
$\twoParticleResistanceNorm_{12}^{xx}$ and
$\twoParticleResistanceNorm_{12}^{yy}$ for $\PPdistance/d\gg1$ are
opposite, consistent with expressions \refeq{far-field flow produced
by a moving particle} and \refeq{far field pressure}.

\subsection{Linear arrays of spheres}
\label{Linear arrays of shperes}

In order to illustrate the role of the far-field flow for in
wall-bounded systems, we present, in Fig.~\ref{new polymer gap}, the
resistance function of rigid linear arrays of $N$ touching spheres.
The spheres are placed in the mid-plane between the walls on a line
pointing in the $x$ direction.  The figure shows the diagonal
components of the translational resistance matrix of the array treated
as a single rigid body
\begin{equation}
\label{rigid-body translational resistance}
\rigidBodyResistanceNorm^{\alpha\alpha}
   =(N\StokesResistance)^{-1}
    \sum_{i,j=1}^N
      \resistanceMatrixElement_{ij}^{\transl\transl\,\alpha\alpha},
\qquad\alpha=x,y,z.
\end{equation}
The normalization of the resistance matrix \refeq{rigid-body
translational resistance} corresponds to the hydrodynamic friction
evaluated per one sphere.  The results shown in the Fig.~\ref{new
polymer gap} are further rescaled by the corresponding one particle
results \refeq{normalized lateral and vertical single-particle
friction}, and they are plotted versus the normalized wall separation
$H/d$ for several values of the chain length $N$.

The results in Fig.~\ref{new polymer gap} indicate that for large
separations between the walls (compared to the chain length) all three
components of the resistance matrix
$\rigidBodyResistanceNorm^{\alpha\alpha}$ decrease monotonically with
$N$.  Consistent with the behavior of elongated particles in unbounded
space \cite{Weinberger:1972,Blawzdziewicz-Wajnryb-Given-Hubbard:2005},
we find that $\rigidBodyResistanceNorm^{\alpha\alpha}\sim 1/\log N$
and $\rigidBodyResistanceNorm^{yy}\simeq\rigidBodyResistanceNorm^{zz}
\simeq2\rigidBodyResistanceNorm^{xx}$ for $1\ll N\ll H/2a$.  In
contrast, for moderate and small values of the wall separation $H$ the
behavior of each component $\rigidBodyResistanceNorm^{\alpha\alpha}$
of the resistance matrix \refeq{rigid-body translational resistance}
is qualitatively different.  The longitudinal component
$\rigidBodyResistanceNorm^{xx}$ decreases monotonically with $N$,
while the other two components $\rigidBodyResistanceNorm^{yy}$ and
$\rigidBodyResistanceNorm^{zz}$ increase with $N$ due to backflow
associated with the presence of the walls.  This effect is
particularly pronounced for the transverse component
$\rigidBodyResistanceNorm^{yy}$, where the increase is by a factor
greater than three for $N=20$ in the regime $H/d\approx 1.2$.

The qualitatively different behavior of the transverse resistance
coefficients $\rigidBodyResistanceNorm^{xx}$ and
$\rigidBodyResistanceNorm^{yy}$ is associated with the opposite
directions of the asymptotic flow field \refeq{far-field flow produced
by a moving particle} on the horizontal lines parallel and
perpendicular to the velocity of a particle.  According to relations
\refeq{far-field flow produced by a moving particle} and \refeq{far
field pressure}, the flow field $\bv^{\rm as}$ in front and behind the
moving particle points in the direction of the particle velocity.
This results in a cooperative effect leading to a reduced resistance
per particle for the longitudinal motion of an array.  The direction
of the flow on the perpendicular line is opposite, which produced a
cumulative effect leading to a large increase of the resistance
coefficient $\rigidBodyResistanceNorm^{yy}$.  This effect is further
discussed in Ref.~\cite{Bhattacharya-Blawzdziewicz-Wajnryb:2005}.

\subsection{Superposition approximation}
\label{Superposition approximation}

To illustrate the effect of the hydrodynamic interactions between
walls on particle dynamics in wall-bounded systems, the results of our
accurate numerical calculations are compared to the single-wall
superposition approximation
\cite{Lobry-Ostrowsky:1996,Jones:2004,Pesche-Nagele:2000}
\begin{equation}
\label{wall superposition for multi-particle system}
\resistanceMatrixSupWNP_{ij}=
   \resistanceMatrixLNP_{ij}+\resistanceMatrixUNP_{ij}
                       -\resistanceMatrixNoWall_{ij}.
\end{equation}
In the above equation, $\resistanceMatrixLNP_{ij}$
($\resistanceMatrixUNP_{ij}$) represents the friction matrix for a
system of $N$ particles in the presence of the lower (upper) wall, and
$\resistanceMatrixNoWall_{ij}$ is the corresponding friction matrix in
the absence of the walls.  We emphasize that, unlike the superposition
terms in equations \refeq{lubrication superposition for resistance
matrix}-\refeq{particle-wall superposition}, all quantities on the
right side of equation \refeq{wall superposition for multi-particle
system} represent  the full $N$-particle friction matrices---the
superposition refers only to the wall contributions.  The subtraction
of the free-space term $\resistanceMatrixNoWall_{ij}$ assures that the
matrix $\resistanceMatrixSupWNP_{ij}$ has a correct limit if the
distance of the particles to one of the walls (or both walls) tends to
infinity.

In Fig.~\ref{new two particles}, the translational friction
coefficients
\begin{equation}
\label{notational keybord twister}
\resistanceMatrixElementSuppNormWNPTT{\alpha\alpha}_{ij}=
\resistanceMatrixElementSuppWNPTT{\alpha\alpha}_{ij}/\StokesResistance,
\end{equation}
evaluated in the superposition approximation, are plotted along with
the accurate results for the two particle system.  The right panel of
Fig.~\ref{new single particle} represents the ratio
$\twoParticleResistanceNorm_{11}^{\alpha\alpha}/
\resistanceMatrixElementSuppNormWNPTT{\alpha\alpha}_{11}$ for a single
particle.  The results indicate that the superposition approximation
is quite accurate for the single-particle friction coefficients and
the self-components of the two-particle friction matrix---the maximal
error for these quantities is about 18\% (also see the more extensive
one-particle calculations reported in Ref.~\cite{Jones:2004}).  

The superposition approximation is much less accurate for the mutual
components $\twoParticleResistanceNorm_{12}^{\alpha\alpha}$ of the
two-particle friction matrix, as shown in the right panels of
Fig.~\ref{new two particles}.  The accuracy of approximation
\refeq{wall superposition for multi-particle system} is especially
poor for the transverse component
$\twoParticleResistanceNorm_{12}^{\alpha\alpha}$ for small values of
the wall separation $H$.  Moreover, the approximation yields an
incorrect $O(\rho^{-1})$ asymptotic behavior of the mutual friction
coefficients for large $\rho$.

The failure of the superposition approximation is particularly
pronounced for the transverse component of the friction coefficient
\refeq{rigid-body translational resistance} for rigid chains of
spheres.  As shown in the second panel of Fig.~\ref{Super Accuracy},
relation \refeq{wall superposition for multi-particle system} grossly
underestimates the coefficient $\rigidBodyResistanceNorm^{yy}$ for
long chains, especially for small and moderate values of the
normalized wall separation $H/d$. The superposition approximation is
insufficient in this regime, because it does not accurately reproduce
the far-field interparticle interactions associated with the flow
\refeq{far-field flow produced by a moving particle}.

\section{Conclusions}
\label{Conclusions}

This paper presents results of a theoretical and numerical study of
many-body hydrodynamic interactions in suspensions of spherical
particles confined between two parallel planar walls.  Our primary
results include the derivation of transformation relations between
spherical and Cartesian basis sets of solutions of Stokes equations.
The transformation formulas enable construction of Stokes-flow fields
that satisfy appropriate boundary conditions both on the planar walls
and on the spherical particle surfaces. Using these transformations,
we have developed an efficient numerical procedure for evaluating
the many-body resistance matrix characterizing hydrodynamic forces
acting on suspension particles in the two-wall geometry.

The basis sets of Stokes flows that are employed in our analysis are
closely related to the spherical solutions introduced by Lamb
\cite{Lamb:1945} and Cartesian solutions introduced by Faxen
\cite{Faxen:1923} (See also section 7.4 of
\cite{Happel-Brenner:1986}).  By a careful choice of the defining
properties, however, we have achieved a symmetric matrix formulation
of the hydrodynamic-interactions problem.  The underlying symmetries
of the basis sets include the curl expressions linking the basis
fields of different tensorial character, and the diagonal
representations of the Oseen tensor in the spherical and Cartesian
bases.  Exploring the symmetry relations in our canonical formulation,
the problem has been reduced to a set of simple explicit expressions.

The results of our theoretical analysis were implemented numerically
in an algorithm for evaluating the many-particle resistance matrix in
the two-wall system.  As a whole, the algorithm is quite complex,
because it involves a large number of components.  These components
include constructing matrix elements of the Green function in terms of
lateral Fourier integrals, using subtraction techniques to improve
convergence of the integrals for configurations with widely separated
particles, solving a linear system of equations for induced-force
multipoles, and correcting the solution for slowly convergent
lubrication contributions.  All the elements in the procedure,
however, are either given explicitly or in terms of simple
quadratures.  

Our numerical algorithm has been used to evaluate the hydrodynamic
resistance matrix for a single particle, a pair of particles, and
linear arrays of particles confined between two planar walls.  The
results for the linear arrays indicate that the far-field flow in
many-particle systems may produce significant collective effects.  A
characteristic example is the large hydrodynamic resistance for the
transverse motion of an elongated array in a narrow space between the
walls.  A simple superposition approximation in which the flow
scattered from the walls is represented as a combination of two
single-wall contributions fails to describe such collective phenomena.

Our current implementation of the Stokesian-dynamics algorithm for
suspensions confined between two parallel planar walls allows
evaluation of hydrodynamic interactions in a system of about a hundred
particles.  For a given number of particles, the numerical cost of the
method increases with the particle separation (especially for $\rho>
20 H$).  This increase results from the oscillatory character of the
integrands in the Fourier representation of the matrix elements of the
Oseen integral operator.  The limitation can be removed by subtracting
several terms of the multiple-image sequence for the flow produced by
the force multipoles induced on the particles.  The subtracted
contributions can be evaluated explicitly
\cite{Bhattacharya-Blawzdziewicz:2002}.

An alternative and more efficient approach is to use asymptotic
expressions for the far-field form of the flow field produced by the
force multipoles in the space between the walls.  We have recently
derived a complete set of such expressions.  An important advantage of
this approach is its simplicity---the asymptotic multipolar flow
fields can be obtained from the solution of the two-dimensional
Laplace's equation for the pressure field in the Hele-Shaw
approximation.  In particular, algorithms based on this method can be
relatively easy generalized for periodic systems.  Moreover, the
efficiency of such algorithms can be substantially improved by
applying the acceleration methods that have been developed for
Laplace's equation \cite{Frenkel-Smit:2002}.  We will describe these
results in forthcoming publications.

S.\,B.\ would like to acknowledge the support by NSF grant
CTS-0201131.  E.\,W.\ was supported by NASA grant NAG3-2704 and in
part by KBN grant No.\ 5T07C 035 22.  J.\,B.  was supported in part by
NSF grant CTS-S0348175 and in part by Hellman Foundation.

\appendix

\section{Spherical basis}
\label{Spherical basis}

In this appendix we list expressions for the reciprocal basis sets
\refeq{spherical basis v -}, \refeq{spherical basis v +} and
\refeq{spherical basis w -}, \refeq{spherical basis w +} in terms of
the normalized vector spherical harmonics, as defined by Edmonds
\cite{Edmonds:1960},
\begin{subequations}
\label{v-harmonics}
\begin{equation}
\label{vector harmonics A}
  {\bf Y}_{ll-1m}(\hr)=\alpha_l^{-1}
  r^{-l+1}\bnabla\left[r^l Y_{lm}(\hr)\right],
\end{equation}
\begin{equation}
\label{vector harmonics B}
  {\bf Y}_{ll+1m}(\hr)=\beta_l^{-1}
  r^{l+2}\bnabla\left[r^{-(l+1)} Y_{lm}(\hr)\right],
\end{equation}
\begin{equation}
\label{vector harmonics C}
  {\bf Y}_{llm}(\hr)=\gamma_l^{-1}\br\times\bnabla_{\mathrm{s}} Y_{lm}(\hr).
\end{equation}
\end{subequations}
Here
\begin{equation}
\label{scalar harmonics}
   Y_{lm}(\hr) =n_{lm}^{-1} (-1)^m P_l^m(\cos\theta)e^{{\rm i}m\varphi}
\end{equation}
are the normalized scalar spherical harmonics, and the normalization
coefficients are
\begin{subequations}
\label{norm.const.}
\begin{equation}
\label{normalization constants a}
  \alpha_l=[l(2l+1)]^{1/2},
\end{equation}
\begin{equation}
\label{normalization constants b}
  \beta_l=[(l+1)(2l+1)]^{1/2},
\end{equation}
\begin{equation}
\label{normalization constants c}
  \gamma_l=-\im[l(l+1)]^{1/2},
\end{equation}
\end{subequations}
and
\begin{equation}
\label{normalization coefficients}
n_{lm}=\left[\frac{4\pi}{2l+1} \frac{(l+m)!}{(l-m)!}\right]^{1/2}.
\end{equation} 
The vector spherical harmonics \refeq{v-harmonics} obey the
orthogonality relations
\begin{equation}
\label{orthogonality relation for Y}
\langle\delta(a){\bf Y}_{lnm}\mid{\bf Y}_{l'n'm'}\rangle
   =\delta_{ll'}\delta_{nn'}\delta_{mm'}.
\end{equation} 

The angular functions $\sphericalBasisCoefPM{lm\sigma}$ and
$\reciprocalSphericalBasisCoefPM{lm\sigma}$ in equations
\refeq{spherical basis v -}, \refeq{spherical basis v +} and
\refeq{spherical basis w -}, \refeq{spherical basis w +} have the
following spherical-harmonics expansions
\begin{subequations}
\label{spherical harmonics expansions of V and W}
\begin{equation}
\label{spherical harmonics expansion of V}
\sphericalBasisCoefPM{lm\sigma}
   =\sum_{\sigma'}\bY_{l\,l-1+\sigma'\,m}V^\pm(l;\sigma'\mid\sigma),
\end{equation}
\begin{equation}
\label{spherical harmonics expansion of W}
\reciprocalSphericalBasisCoefPM{lm\sigma}
   =\sum_{\sigma'}\bY_{l\,l-1+\sigma'\,m}W^\pm(l;\sigma'\mid\sigma),
\end{equation}
\end{subequations}
The explicit expressions for the matrices ${\sf V}^\pm$ at the right
side of \refeq{spherical harmonics expansions of V and W} are
\begin{equation}
\label{matrix V^+}
{\sf V}^{+}(l)=
\left[\begin{array}{ccc}
\alpha_l &  0  & \frac{l}{2(2l+1)}\alpha_l\\
0 & \frac{i}{l+1}\gamma_l & 0\\
0 & 0 & \frac{l}{(l+1)(2l+1)(2l+3)}\beta_l
\end{array}
\right]
\end{equation}
and
\begin{equation}
\label{matrix V^-}
{\sf V}^{-}(l)=\frac{1}{2l+1}
\left[\begin{array}{ccc}
\frac{l+1}{l(2l-1)(2l+1)}\alpha_l & 0 & 0\\
0 & \im l^{-1}\gamma_l & 0\\
-\frac{1}{2(2l+1)}\beta_l & 0 & \beta_l
\end{array}
\right].
\end{equation}
Due to orthogonality relations \refeq{orthogonality relations for
spherical reciprocal basis} and \refeq{orthogonality relation for Y},
the matrices ${\sf W}^\pm$ and ${\sf V}^\pm$ satisfy the corresponding
orthogonality condition 
\begin{equation}
\label{vw}
[{\sf W}^\pm]^{\dagger}=[{\sf V}^\pm]^{-1},
\end{equation}
which yields
\begin{equation}
\label{matrix W^+}
{\sf W}^+(l)=
\left[\begin{array}{ccc}
\alpha_l^{-1} & 0 & 0\\
0 & -i(l+1)\gamma_l^{-1} & 0\\
-\frac{(l+1)(2l+3)}{2l}\beta_l^{-1} & 0 & 
   \frac{(l+1)(2l+1)(2l+3)}{l}\beta_l^{-1}
\end{array}
\right],
\end{equation}
\begin{equation}
\label{matrix W^-}
{\sf W}^-(l)=
(2l+1)\left[\begin{array}{ccc}
\frac{l(2l-1)(2l+1)}{l+1}\alpha_l^{-1} & 0 & 
   \frac{l(2l-1)}{2(l+1)}\alpha_l^{-1}\\
0 & -il\gamma_l^{-1} & 0\\
0 & 0 & \beta_l^{-1}.
\end{array}
\right].
\end{equation}

In the original publication \cite{Cichocki-Felderhof-Schmitz:1988} and
in following papers
\cite{
Cichocki-Jones-Kutteh-Wajnryb:2000,%
Blawzdziewicz-Vlahovska-Loewenberg:2000%
}, 
the basis functions $\sphericalBasisPM{lm\sigma}$ and
$\reciprocalSphericalBasisPM{lm\sigma}$ were normalized differently.
The relation between the spherical basis fields $\bv^{\pm({\rm
CFS})}_{lm\sigma}$ and $\bw^{\pm({\rm CFS})}_{lm\sigma},$ in the
original normalization of Cichocki
\etal~\cite{Cichocki-Felderhof-Schmitz:1988} and the basis introduced
in the present paper is
\begin{subequations}
\label{relation between BBW and CFS bases}
\begin{equation}
\label{relation between BBW and CFS v bases}
\sphericalBasisM{lm\sigma}(\br)
   =N_{l\sigma}^{-1}n_{lm}^{-1}\bv^{-({\rm CFS})}_{lm\sigma}(\br),
\qquad
\sphericalBasisP{lm\sigma}(\br)
   =N_{l\sigma}n_{lm}^{-1}\bv^{+({\rm CFS})}_{lm\sigma}(\br),
\end{equation}
\begin{equation}
\label{relation between BBW and CFS w bases}
\reciprocalSphericalBasisM{lm\sigma}(\br)
   =N_{l\sigma}n_{lm}r\bw^{-({\rm CFS})}_{lm\sigma}(\br),
\qquad
\reciprocalSphericalBasisP{lm\sigma}(\br)
   =N_{l\sigma}^{-1}n_{lm}r\bw^{+({\rm CFS})}_{lm\sigma}(\br),
\end{equation}
\end{subequations}
where
\begin{equation}
\label{change-of-normalization coefficients}
N_{l0}=1,\qquad N_{l1}=-{(l+1)^{-1}},\qquad l[(l+1)(2l+1)(2l+3)]^{-1}.
\end{equation}

\section{Transformation vectors 
$\frictionProjectionVector^\transl$ and
$\frictionProjectionVector^\rot$}
\label{Transformation vectors X}

The transformation vectors
$\tilde\frictionProjectionVector^\transl(m)$ and
$\tilde\frictionProjectionVector^\rot(m)$, $m=-1,0,1$ in relations
\refeq{friction projection matrix} are obtained by inserting the
multipolar expansion \refeq{induced force in terms of multipoles} into
the definitions \refeq{force and torque} and \refeq{force and torque
projections}.  The resulting expressions are evaluated using formulas
\refeq{spherical basis w +}, \refeq{spherical harmonics expansion of
W}, and \refeq{matrix W^+}, which yield
\begin{equation}
\label{friction projection vector translation}
\tilde\frictionProjectionVector^\transl(-1)=
({\textstyle\frac{2}{3}}\upi)^{1/2}
   \left[
      \begin{array}{c}
         1\\-\im\\0
      \end{array}
   \right],
\quad
\tilde\frictionProjectionVector^\transl(0)=
({\textstyle\frac{2}{3}}\upi)^{1/2}
   \left[
      \begin{array}{c}
         0\\0\\\sqrt{2}
      \end{array}
   \right],
\quad
\tilde\frictionProjectionVector^\transl(1)=
({\textstyle\frac{2}{3}}\upi)^{1/2}
   \left[
      \begin{array}{c}
         -1\\-\im\\0
      \end{array}
   \right],
\end{equation}
and
\begin{equation}
\label{relation between X r and x t}
\tilde\frictionProjectionVector^\rot(m)=-2\im 
\tilde\frictionProjectionVector^\transl(m),\qquad m=-1,0,1.
\end{equation}

\section{Elements of transformation matrices 
$\tildeTransformationSC{+\pm}$ and $\tildeTransformationCS{\pm-}$}
\label{Derivation of coefficients a,b,c}

Due to the symmetric formulation of the problem, all four
transformation matrices \refeq{form of transformation SC and CS}
depend on the same set of coefficients \refeq{coefficients a,b,c}.
Thus, it is sufficient to derive the explicit expression
for only one of these matrices.  Here we focus on the transformation
relation \refeq{Cartesian - in spherical -- same point} between the
Cartesian and spherical basis fields $\CartesianBasisP{\bk\sigma}$ and
$\sphericalBasisP{lm\sigma}$.

To find the required transformation formula, we expand
$\CartesianBasisP{\bk\sigma}$ in powers of the radial coordinate $r$.
The expansion can be represented in the form
\begin{equation}
\label{expansion of Cartesian v+ in powers of r}
\CartesianBasisP{\bk\sigma}(\br)=\sum_{n=1}^\infty 
   \CartesianBasisPn{\bk\sigma}{n}(\br),
\end{equation}
where the fields $\CartesianBasisPn{\bk\sigma}{n}(\br)$ are
homogeneous functions of order $n$ in $r$.  We note that each term in
the expansion \refeq{expansion of Cartesian v+ in powers of r} is
itself a Stokes flow, because the linear operators in the Stokes
equations do not couple terms with different powers of $r$.  It
follows that the consecutive expansion terms can be represented as
combinations of the spherical basis solutions \refeq{spherical basis v
+} with $l+\sigma-1=n$.  For the pressure solution
$\CartesianBasisP{\bk2}$, this representation can be expressed as
\begin{equation}
\label{triangular form of expansion of Cartesian v+ 2}
\CartesianBasisPn{\bk2}{n}=\bu^{(n)}_{\bk2}+\bu^{(n)}_{\bk1}+\bu^{(n)}_{\bk0},
\end{equation}
where
\begin{equation}
\label{expansion of n terms in spherical basis}
\bu^{(l+\sigma-1)}_{\bk\sigma}
   =\sum_{m=-l}^l a^{lm}_{k\sigma}\sphericalBasisP{lm\sigma}.
\end{equation}
Comparing the above expressions with equations \refeq{Cartesian - in
spherical -- same point}, \refeq{matrix K}, and \refeq{form of
transformation SC} yields
\begin{equation}
\label{relation between a l_m_sigma and a_sigma}
a^{lm}_{k\sigma}
   =\im^m(2\upi k)^{-1/2}k^{l+\sigma+1}\e^{-\im m\psi}\bar a_\sigma,
\end{equation}
where the coefficients $a_\sigma$ correspond to the coefficients
$a,b,c$ in equation \refeq{form of transformation SC},
\begin{equation}
\label{relation between a_sigma and a,b,c}
\bar a_0\equiv c,\qquad \bar a_1\equiv 2b,\qquad \bar a_2\equiv 4a.
\end{equation}

The remaining components $\CartesianBasisPn{\bk1}{n}$ and
$\CartesianBasisPn{\bk0}{n}$ can be related to the flow fields
\refeq{expansion of n terms in spherical basis} by applying the curl
operator to both sides of equation \refeq{expansion of Cartesian v+ in
powers of r} with $\sigma=2$ and $\sigma=1$.  After inserting
decomposition \refeq{triangular form of expansion of Cartesian v+ 2}
and using curl relations \refeq{Cartesian curl +} and (\ref{spherical
zero curl}\textit{b}), we collect terms corresponding to the same
power of $r$, which yields
\begin{subequations}
\label{triangular form of expansion of Cartesian v+}
\begin{equation}
\label{triangular form of expansion of Cartesian v+ 1}
   \CartesianBasisPn{\bk1}{n-1}=\half\im k^{-1}\bnabla\btimes
\left(
   \bu^{(n)}_{\bk2}+\bu^{(n)}_{\bk1}
\right),
\end{equation}
\begin{equation}
\label{triangular form of expansion of Cartesian v+ 0}
\CartesianBasisPn{\bk0}{n-2}
   =-\quarter k^{-2}\bnabla\btimes
      \left(\bnabla\btimes\bu^{(n)}_{\bk2}\right).
\end{equation}
\end{subequations}

The above results are consistent with the triangular structure of the
transformation matrix \refeq{form of transformation SC and CS}.  To
evaluate the coefficients $a^{lm}_{k\sigma}$, we reduce expressions \refeq{triangular form of expansion
of Cartesian v+ 2} and \refeq{triangular form of expansion of
Cartesian v+} to equivalent relations between appropriately defined
harmonic scalar fields.   For the spherical basis flows
$\sphericalBasisP{lm\sigma}$ we have
\begin{equation}
\label{spherical basis + in terms of scalar solid harmonics}
\sphericalBasisP{lm\sigma}=\uncurlS_{l\sigma}\varPhi^+_{lm},
\end{equation}
where
\begin{equation}
\label{scalar solid harmonic}
\varPhi^+_{lm}(\br)=r^lY_{lm}(\hat\br),
\end{equation}
and the operators $\uncurlS_{l\sigma}$ are given by
\begin{subequations}
\label{un-curl operators spherical}
\begin{equation}
\label{un-curl operators spherical 0}
\uncurlS_{l0}=\bnabla,
\end{equation}
\begin{equation}
\label{un-curl operators spherical 1}
\uncurlS_{l1}=\im(l+1)^{-1}\br\btimes\bnabla,
\end{equation}
\begin{equation}
\label{un-curl operators spherical 2}
\uncurlS_{l2}=[(l+1)(2l+3)]^{-1}
   [
      -l\br
      +\half(l+3)r^2\bnabla
   ].
\end{equation}
\end{subequations}
The analogous expressions for the Cartesian basis fields
$\CartesianBasisP{\bk\sigma}$ are
\begin{equation}
\label{Cartesian basis + in terms of scalar solid harmonics}
\CartesianBasisP{\bk\sigma}=\uncurlC_{k\sigma}\varPhi^+_{\bk},
\end{equation}
where
\begin{equation}
\label{scalar Cartesian harmonic}
\varPhi^+_{\bk}(\br)=(32\upi^2k)^{-1/2}\e^{\im\bk\bcdot\brho+kz}.
\end{equation}
The operators $\uncurlC_{k0}$ and $\uncurlC_{k1}$ are given by the
expressions
\begin{subequations}
\label{un-curl operators Cartesian}
\begin{equation}
\label{un-curl operators Cartesian 0}
\uncurlC_{k0}=k^{-1}\bnabla,
\end{equation}
\begin{equation}
\label{un-curl operators Cartesian 1}
\uncurlC_{k1}=2ik^{-1}\ez\btimes\bnabla,
\end{equation}
and the operator $\uncurlC_{k2}$ is given by
\begin{equation}
\label{homogeneous decomposition of un-curl C 2}
\uncurlC_{k2}=\uncurlCA{k}+\uncurlCB{k},
\end{equation}
where
\begin{equation}
\label{homogeneous parts of un-curl C 2}
   \uncurlCA{k}=k^{-1}\bnabla,
\qquad
\uncurlCB{k}=
   -2\ez+2z\bnabla.
\end{equation}
\end{subequations}
The above relations can easily be verified using expressions
\refeq{spherical basis v +} and \refeq{matrix V^+} for the spherical
basis fields and expressions \refeq{Cartesian basis +} for the
Cartesian basis.

The radial-expansion components $\CartesianBasisPn{\bk2}{n}$ in the
decomposition \refeq{expansion of Cartesian v+ in powers of r} of the
Cartesian basis flows can be determined by applying the operators
$\uncurlC_{k\sigma}$ to the expansion of the Cartesian scalar field
\refeq{scalar Cartesian harmonic} in powers of $r$
\begin{equation}
\label{Cart. Scalar expansion}
\varPhi^+_{\bk}=\sum_{n=0}^{\infty}\varPhi^{+(n)}_{\bk},
\end{equation}
where
\begin{equation}
\label{Phi^(n)}
\varPhi^{+(n)}_{\bk}(\br)
   =(32\upi^2k)^{-1/2}\frac{(\im\bk\bcdot\brho+kz)^n}{n!}.
\end{equation}
Inserting relations \refeq{Cartesian basis + in terms of scalar solid
harmonics} and \refeq{Cart. Scalar expansion} into the expansion
\refeq{expansion of Cartesian v+ in powers of r} and collecting terms
corresponding to a given power of $r$ we find
\refstepcounter{equation}
$$
\label{L expressions for v^(n) sigma=0,1}
\CartesianBasisPn{\bk0}{n}=\uncurlC_{k0}\varPhi^{+(n+1)}_{\bk},
\qquad
\CartesianBasisPn{\bk1}{n}=\uncurlC_{k1}\varPhi^{+(n+1)}_{\bk},
\eqno{(\theequation{\mathit{a},\mathit{b}})}
$$
and
$$
\label{L expressions for v^(n) sigma=2}
\CartesianBasisPn{\bk2}{n}
   =\uncurlCA{k}\varPhi^{+(n+1)}_{\bk}
   +\uncurlCB{k}\varPhi^{+(n)}_{\bk}.
\eqno{(\theequation{\mathit{c}})}
$$
With the help of relations \refeq{spherical basis + in terms of scalar
solid harmonics}--\refeq{un-curl operators spherical} for the spherical
basis fields, the flow fields $\bu^{(n)}_{\bk\sigma}$ in equations
\refeq{triangular form of expansion of Cartesian v+} can be
represented in a similar manner,
\begin{equation}
\label{L expressions for u^(n)}
\bu^{(l+\sigma-1)}_{\bk\sigma}=\uncurlS_{l\sigma}\varPsi^{+(l)}_{\bk\sigma},
   \qquad\sigma=0,1,2,
\end{equation}
where
\begin{equation}
\label{Psi^(l)}
\varPsi^{+(l)}_{\bk\sigma}=\sum_{m=-l}^l a^{lm}_{k\sigma}\varPhi^{+}_{lm},
\end{equation}
according to equation \refeq{expansion of n terms in spherical basis}.

A closed set of equations for the scalar functions
$\varPsi^{+(l)}_{\bk\sigma}$ is obtained by inserting relation
\refeq{L expressions for u^(n)} into \refeq{triangular form of
expansion of Cartesian v+ 2} and \refeq{triangular form of expansion
of Cartesian v+}, using \refeq{L expressions for v^(n) sigma=0,1}, and
employing the curl identities \refstepcounter{equation}
$$
\label{curl uncurl identities spherical}
\im\bnabla\btimes\uncurlS_{l2}\varPsi_l
   =\uncurlS_{l1}\varPsi_l,
\qquad
\im\bnabla\btimes\uncurlS_{l1}\varPsi_l
   =\uncurlS_{l0}\varPsi_l
\eqno{(\theequation{\mathit{a},\mathit{b}})},
$$ 
where $\varPsi_l$ is an arbitrary solid harmonic of the order $l$.
The above expressions correspond to the curl identities
\refeq{spherical curl +} for the spherical basis fields, and can be
verified using relations \refeq{un-curl operators spherical}.  The
equations for the scalar functions $\varPsi^{+(l)}_{\bk\sigma}$
derived by this procedure are
\begin{subequations}
\label{equations for Psi}
\begin{equation}
\label{equation for Psi 2}
\uncurlS_{l0}\varPsi^{+(l)}_{\bk2}
   =4k^2\uncurlC_{k0}\varPhi^{+(l)}_{\bk},
\end{equation}
\begin{equation}
\label{equation for Psi 1}
\uncurlS_{l0}\varPsi^{+(l)}_{\bk1}
   =2k\uncurlC_{k1}\varPhi^{+(l)}_{\bk}
   -\uncurlS_{l-11}\varPsi^{+(l-1)}_{\bk2},
\end{equation}
\begin{equation}
\label{equation for Psi 0}
\uncurlS_{l0}\varPsi^{+(l)}_{\bk0}
   =\uncurlCA{k}\varPhi^{+(l)}_{\bk}
      +\uncurlCB{k}\varPhi^{+(l-1)}_{\bk}
         -\uncurlS_{l-11}\varPsi^{+(l-1)}_{\bk1}
            -\uncurlS_{l-22}\varPsi^{+(l-2)}_{\bk2}.
\end{equation}
\end{subequations}

The above equations can be explicitly solved for the unknown fields
$\varPsi^{+(l)}_{\bk\sigma}$.  Using expressions \refeq{un-curl
operators spherical} and \refeq{un-curl operators Cartesian} for the
operators $\uncurlS_{l\sigma}$ and $\uncurlC_{k\sigma}$, and
simplifying the results using relation \refeq{Phi^(n)} for the
field $\varPhi^{+(l)}_{\bk}$ we find
\begin{subequations}
\label{solutions for Psi}
\begin{equation}
\label{solution for Psi 2}
\varPsi^{+(l)}_{\bk2}=4k\varPhi^{+(l)}_{\bk},
\end{equation}
\begin{equation}
\label{solution for Psi 1}
\varPsi^{+(l)}_{\bk1}=-\frac{4ky}{l}\varPhi^{+(l-1)}_{\bk},
\end{equation}
and
\begin{equation}
\label{solution for Psi 0}
\varPsi^{+(l)}_{\bk0}
   =\frac{k[(2l^2-4l+3)(\im x+z)^2+2(l-2)z(\im x+z)
   -2(l-1)(l-2)y^2]}{l(l-1)(2l-1)}\varPhi^{+(l-2)}_{\bk},
\end{equation}
\end{subequations}
where $\bk=k\ex$ is assumed.

In the final step of our derivation, we recall that the functions
$\varPsi^{+(l)}_{\bk2}$ are solid harmonics of order $l$, according to
expressions \refeq{scalar solid harmonic} and \refeq{Psi^(l)}.  To
obtain the expansion coefficients $a^{lm}_{k\sigma}$ in relation
\refeq{Psi^(l)} for even values of the parameter $l+m$, we evaluate
both sides of \refeq{solutions for Psi} on the plane $z=0$ and compare
the coefficients of the angular Fourier modes $\e^{\im m\phi}$.  In
the case of odd values of the parameter $l+m$, a similar analysis is
performed for the derivative of both sides of equations
\refeq{solutions for Psi} with respect to the coordinate $z$.  The
analysis yields the quantities $a^{lm}_{k\sigma}$ in the form
\refeq{relation between a l_m_sigma and a_sigma}, with the
coefficients \refeq{relation between a_sigma and a,b,c} given by
expressions \refeq{coefficients a,b,c}.

\section{Large $k$ behavior of integrands $\delta\Psi(k)$}
\label{Large k behavior of integrands delta Psi}

In this appendix we derive the asymptotic expression \refeq{asymptotic
behavior of wall-interaction integrand} for the large $k$ behavior of
integrand $\delta\Psi(k)$.  According to equations \refeq{expression
for g two wall} and \refeq{two wall integrand}, the decomposition
\refeq{decomposition of integrand} of the integrand $\Psi(k)$
corresponds to the separation
\begin{equation}
\label{decomposition of two-wall Z matrix}
\tildeZW(k)=\tildeZWzero+\delta\tildeZW(k)
\end{equation}
of the two-wall scattering matrix \refeq{two wall Z matrix scaled}
into the $O(1)$ diagonal contribution
\begin{equation}
\label{diagonal contribution to two-wall Z}
\tildeZWzero=
\left[
   \begin{array}{cc}
      \ZsingleWall&0
\\\\
      0&\ZsingleWall
   \end{array}
\right]
\end{equation}
and the correction of the form
\begin{equation}
\label{correction to two-wall Z matrix}
\delta\tildeZW(k)=
   -\tildeZWzero\bcdot\tildeSTW(k)\bcdot\tildeZW(k),
\end{equation}
where
\begin{equation}
\label{off-diagonal contribution to two-wall Z}
\tildeSTW(k)=
\left[
   \begin{array}{cc}
      0&\tildeCartesianDisplacement{++}(-kH)
\\\\
      \tildeCartesianDisplacement{--}(kH)&0
   \end{array}
\right].
\end{equation}
Taking into account relation \refeq{expression for tilde S} we find
that
\begin{equation}
\label{order of correction term in two-wall Z}
\delta\tildeZW(k)\sim\e^{-kH},
\end{equation}
which implies that
\begin{equation}
\label{correction smaller than Z zero}
\delta\tildeZW(k)\ll\tildeZWzero, \qquad k\gg1.
\end{equation}

Inserting the decomposition \refeq{decomposition of two-wall Z matrix}
into equations \refeq{expression for g two wall} and \refeq{two wall
integrand} yields
\begin{equation}
\label{correction to two wall integrand}
\delta\integrandTwoWall(k)=
      \delta\TwoWallsGreenFourierElement(
         k;lm\sigma\mid l'm'\sigma')
       k^{l+l'+\sigma+\sigma'-2}\BesselJ_{m'-m}(k\rho_{ij}),
\end{equation}
where
\begin{equation}
\label{expression for g two wall correction}
\delta\TwoWallsGreenFourier(k;lm\mid l'm')=
   -\tildeTSC(lm)
\bcdot
   \tildeSpw{i}(k)
\bcdot
   \delta\tildeZW(k)
\bcdot
   \tildeSwp{j}(k)
\bcdot
   \tildeTCS(l'm').
\end{equation}
The asymptotic expression \refeq{asymptotic behavior of
wall-interaction integrand} is obtained by inserting relation
\refeq{correction to two-wall Z matrix} with $\ZW(k)\simeq\ZWzero$
into \refeq{expression for g two wall correction}, and using
equations \refeq{factorization of Cartesian displacement matrix},
\refeq{expression for tilde S} and \refeq{two wall displacement}.
Evaluation of the slowest-decaying term yields \refeq{asymptotic
behavior of wall-interaction integrand} with
\begin{equation}
\label{alternative expression for second image offset}
\tilde\imageOffset_{ij}=\min(Z_{iL}+Z_{Uj},Z_{jL}+Z_{Ui})+H,
\end{equation}
which is equivalent to \refeq{distance to second image}.

We note that the convergence of the integrand \refeq{two wall
integrand} can further be improved by subtracting from the two-wall
scattering matrix $\tildeZW$ several terms in the expansion
\begin{equation}
\label{expansion of Z two wall}
   \tildeZW(k)=\sum_{s=0}^\infty(-1)^s
      \tildeZWzero\bcdot
         [\tildeSTW(k)\bcdot\tildeZWzero]^s.
\end{equation}
One can show that the subtracted terms correspond to consecutive
reflections of the flow field from the walls.  Thus, these terms can
be evaluated without numerical integration using the
image-representation formulas derived by two of us
\cite{Bhattacharya-Blawzdziewicz:2002}.

\bibliographystyle{unsrt}
\bibliography{/home/jerzy/BIB/jbib}

\begin{thebibliography}{10}

\bibitem{Lin-Crocker-Prasad-Schofield-Weitz-Lubensky-Yodh:2000}
K.~H. Lin, J.~C. Crocker, V.~Prasad, A.~Schofield, D.~A. Weitz, T.C. Lubensky,
  and A.G. Yodh.
\newblock {Entropically driven colloidal crystalization on patterned surfaces}.
\newblock {\em Phys. Rev. Lett.}, 85:1770, 2000.

\bibitem{Acuna_Campa-Carbajal_Tinoco-Arauz_Lara-Medina_Noyola:1998}
H.~Acu\~{n}a Campa, M.~D. Carbajal-Tinoco, J.~L. Arauz-Lara, and
  M.~Medina-Noyola.
\newblock {Collective dynamics in quasibidimensional colloidal suspensions}.
\newblock {\em Phys. Rev. Lett.}, 80:5802--5, 1998.

\bibitem{Pesche-Kollmann-Nagele:2001}
R.~{Pesch\'{e}}, M.~Kollmann, and G.~N{\"{a}}gele.
\newblock {Brownian dynamics study of dynamic scaling and related freezing
  criteria in quasi-two-dimensional dispersions}.
\newblock {\em J. Chem. Phys.}, 114:8701--7, 2001.

\bibitem{Santana_Solano-Arauz_Lara:2002}
J.~Santana-Solano and J.~L. Arauz-Lara.
\newblock {Short-time dynamics of colloidal particles confined between two
  walls}.
\newblock {\em Phys. Rev. E}, 65:021406--1--8, 2002.

\bibitem{Sethumadhavan-Nikolov-Wasan:2001}
G.~N. Sethumadhavan, A.~D. Nikolov, and D.~T. Wasan.
\newblock {Stability of liquid films containing monodisperse colloidal
  particles}.
\newblock {\em J. Colloid Interface Sci.}, 240:105--12, 2001.

\bibitem{Subramanian-Manoharan-Thorne-Pine:1999}
G.~Subramanian, V.~N. Manoharan, J.~D. Thorne, and D.~J. Pine.
\newblock {Ordered macroporous materials by colloidal assembly: A possible
  route to photonic bandgap materials}.
\newblock {\em Adv. Mater.}, 11:1261--1265, 1999.

\bibitem{Seelig-Tang-Yamilov-Cao-Chang:2002}
E.~W. Seelig, B.~Tang, A.~Yamilov, H.~Cao, and R.~P.~H. Chang.
\newblock {Self-assembled 3D photonic crystals from ZnO colloidal spheres}.
\newblock {\em Mater. Chem. Phys.}, 80:257--63, 2002.

\bibitem{Prieve-Luo-Lanni:1987}
D.~C. Prieve, F.~Luo, and F.~Lanni.
\newblock {Brownian-motion of a hydrosol particle in a colloidal force-field}.
\newblock {\em Faraday Discuss. Chem. Soc.}, 83:297, 1987.

\bibitem{Walz-Suresh:1995}
J.~Y. Walz and L.~Suresh.
\newblock Study of the sedimentation of a single particle toward a flat plate.
\newblock {\em J. Chem. Phys.}, 103:10714--725, 1995.

\bibitem{Faucheux-Libchaber:1994}
L.~P. Faucheux and A.~J. Libchaber.
\newblock {Confined Brownian motion}.
\newblock {\em Phys. Rev. E.}, 49:5158--63, 1994.

\bibitem{Lin-Yu-Rice:2000}
B.~Lin, J.~Yu, and S.~A. Rice.
\newblock {Direct measurements of constrained Brownian motion of an isolated
  sphere between two walls}.
\newblock {\em Phys. Rev. E.}, 62:3909--19, 2000.

\bibitem{Palberg-Biehl:2003}
T.~Palberg and R.~Biehl.
\newblock {Sheared colloidal crystals in confined geometry: a real space study
  on stationary structures under shear}.
\newblock {\em Faraday Discuss.}, 123:133--43, 2003.

\bibitem{Crocker-Matteo-Dinsmore-Yodh:1999}
J.~C. Crocker, J.~A. Matteo, A.~D. Dinsmore, and A.~G. Yodh.
\newblock {Entropic attraction and repulsion in binary colloids probed with a
  line optical tweezer}.
\newblock {\em Phys. Rev. Lett.}, 82:4352--5, 1999.

\bibitem{Kim-Karrila:1991}
S.~Kim and S.~J. Karrila.
\newblock {\em Microhydrodynamics: Principles and Selected Applications}.
\newblock Butterworth-Heinemann, London, 1991.

\bibitem{Durlofsky-Brady-Bossis:1987}
L.~Durlofsky, J.~F. Brady, and G.~Bossis.
\newblock {Dynamic simulation of hydrodynamically interacting particles}.
\newblock {\em J. Fluid Mech.}, 180:21--49, 1987.

\bibitem{Ladd:1988}
A.~J.~C. Ladd.
\newblock {Hydrodynamic interactions in suspensions of spherical particles}.
\newblock {\em J. Chem. Phys.}, 88:5051, 1988.

\bibitem{Cichocki-Felderhof-Hinsen-Wajnryb-Blawzdziewicz:1994}
B.~Cichocki, B.~U. Felderhof, K.~Hinsen, E.~Wajnryb, and J.~B{\l}awzdziewicz.
\newblock {Friction and mobility of many spheres in Stokes flow}.
\newblock {\em J. Chem. Phys.}, 100:3780--3790, 1994.

\bibitem{Sangani-Mo:1996}
A.~S. Sangani and G.~B. Mo.
\newblock {An $O(N)$ algorithm for Stokes and Laplace interactions of
  particles}.
\newblock {\em Phys. Fluids}, 8:1990--2010, 1996.

\bibitem{Sierou-Brady:2001}
A.~Sierou and J.~F. Brady.
\newblock {Accelerated Stokesian dynamics simulations}.
\newblock {\em J. Fluid Mech.}, 448:115--46, 2001.

\bibitem{Cichocki-Jones:1998}
B.~Cichocki and R.~B. Jones.
\newblock Image representation of a spherical particle near a hard wall.
\newblock {\em Physica A}, 258:273--302, 1998.

\bibitem{Cichocki-Jones-Kutteh-Wajnryb:2000}
B.~Cichocki, R.~B. Jones, R.~Kutteh, and E.~Wajnryb.
\newblock {Friction and mobility for colloidal spheres in Stokes flow near a
  boundary: The multipole method and applications}.
\newblock {\em J. Chem. Phys.}, 112:2548--61, 2000.

\bibitem{Lobry-Ostrowsky:1996}
L.~Lobry and N.~Ostrowsky.
\newblock {Diffusion of Brownian particles trapped between two walls: Theory
  and dynamic-light-scattering measurements}.
\newblock {\em Phys. Rev. B}, 53:12050--6, 1996.

\bibitem{Benesch-Yiacoumi-Tsouris:2003}
T.~Benesch, S.~Yiacoumi, and C.~Tsouris.
\newblock {Brownian motion in confinement}.
\newblock {\em Phys. Rev. E}, 68:021401--1--5, 2003.

\bibitem{Ganatos-Weinbaum-Pfeffer:1980}
P.~Ganatos, S.~Weinbaum, and R.~Pfeffer.
\newblock {A strong interaction theory for the creeping motion of a sphere
  between plane parallel boundaries. Part 1. Perpendicular motion}.
\newblock {\em J. Fluid Mech.}, 99:739--53, 1980.

\bibitem{Ganatos-Pfeffer-Weinbaum:1980}
P.~Ganatos, R.~Pfeffer, and S.~Weinbaum.
\newblock {A strong interaction theory for the creeping motion of a sphere
  between plane parallel boundaries. Part 2. Parallel motion}.
\newblock {\em J. Fluid Mech.}, 99:755--83, 1980.

\bibitem{Staben-Zinchenko-Davis:2003}
M.~E. Staben, A.~Z. Zinchenko, and R.~H. Davis.
\newblock {Motion of a particle between two parallel plane walls in
  low-Reynolds-number Poiseuille flow}.
\newblock {\em Phys. Fluids.}, 15:1711--33, 2003.

\bibitem{Bhattacharya-Blawzdziewicz:2002}
S.~Bhattacharya and J.~B{\l}awzdziewicz.
\newblock {Image system for Stokes-flow singularity between two parallel planar
  walls}.
\newblock {\em J. Math. Phys.}, 43:5720--31, 2002.

\bibitem{Durlofsky-Brady:1989}
L.~J. Durlofsky and J.~F. Brady.
\newblock {Dynamic simulation of bounded suspensions of hydrodynamically
  interacting particles}.
\newblock {\em J. Fluid. Mech.}, 200:39--67, 1989.

\bibitem{Nott-Brady:1994}
P.R. Nott and J.F. Brady.
\newblock Pressure-driven flow of suspensions---simulation and theory.
\newblock {\em J. Fluid Mech.}, 275:157--199, 1994.

\bibitem{Morris-Brady:1998}
J.~F. Morris and J.~F. Brady.
\newblock {Pressure-driven flow of a suspension: Buoyancy effects}.
\newblock {\em Int. J. Multiphase Flow}, 24:105--30, 1998.

\bibitem{Jones:2004}
R.~B. Jones.
\newblock {Spherical particle in Poiseuille flow between planar walls}.
\newblock {\em J. Chem. Phys.}, 121:483--500, 2004.

\bibitem{Jones:2004a}
R.~B. Jones.
\newblock {Hydrodynamic interactions of a spherical particle in Poiseuille flow
  between planar walls}.
\newblock {XXI International Congress of Theoretical and Applied Mechanics,
  August 15--21, 2004, Warsaw, Poland}, 2004.

\bibitem{Cox-Brenner:1967}
R.~G. Cox and H.~Brenner.
\newblock {Effect of finite boundaries on Stokes resistance of an arbitrary
  particle .3. translation and rotation}.
\newblock {\em J. Fluid Mech.}, 28:391, 1967.

\bibitem{Mazur-Bedeaux:1974}
P.~Mazur and D.~Bedeaux.
\newblock {A generalization of Fax\'en's theorem to nonsteady motion of a
  sphere through an incompressible fluid in arbitrary flow}.
\newblock {\em Physica}, 76:235--46, 1974.

\bibitem{Felderhof:1976b}
B.~U. Felderhof.
\newblock Force density induced on a sphere in linear hydrodynamics. ii. moving
  sphere, mixed boundary conditions.
\newblock {\em Physica A}, 84:569--576, 1976.

\bibitem{Jones-Schmitz:1988}
R.~B. Jones and R.~Schmitz.
\newblock Mobility matrix for arbitrary spherical particles in solution.
\newblock {\em Physica A}, 149:373--394, 1988.

\bibitem{Cichocki-Felderhof-Schmitz:1988}
B.~Cichocki, B.~U. Felderhof, and R.~Schmitz.
\newblock Hydrodynamic interactions between two spherical particles.
\newblock {\em PhysicoChem. Hyd.}, 10:383--403, 1988.

\bibitem{Blawzdziewicz-Wajnryb-Loewenberg:1999}
J.~B{\l}awzdziewicz, E.~Wajnryb, and M.~Loewenberg.
\newblock Hydrodynamic interactions and collision efficiencies of spherical
  drops covered with an incompressible surfactant film.
\newblock {\em J. Fluid Mech.}, 395:29--59, 1999.

\bibitem{Perkins-Jones:1991}
G.~S. Perkins and R.~B. Jones.
\newblock {Hydrodynamic interaction of a spherical particle with a planar
  boundary. 1. Free-surface}.
\newblock {\em Physica A}, 171:575--604, 1991.

\bibitem{Edmonds:1960}
A.~R. Edmonds.
\newblock {\em Angular Momentum in Quantum Mechanics}.
\newblock Princeton University Press, Princeton, 1960.

\bibitem{Felderhof-Jones:1989}
B.~U. Felderhof and R.~B. Jones.
\newblock {Displacement theorems for spherical solutions of the linear
  Navier-Stokes equations}.
\newblock {\em J. Math. Phys.}, 30:339--42, 1989.

\bibitem{Blawzdziewicz-Cristini-Loewenberg:1999}
J.~B{\l}awz\-dziewicz, V.~Cristini, and M.~Loewenberg.
\newblock Stokes flow in the presence of a planar interface covered with
  incompressible surfactant.
\newblock {\em Phys. Fluids}, 11:251--258, 1999.

\bibitem{Lorentz:1907}
H.~A. Lorentz.
\newblock {A general theory concerning the motion of a viscous fluid}.
\newblock {\em Abhandl. Theor. Phys.}, 1:23, 1907.

\bibitem{Bhattacharya-Blawzdziewicz-Wajnryb:2005}
S.~Bhattacharya, J.~B{\l}awzdziewicz, and E.~Wajnryb.
\newblock {Hydrodynamic interactions of spherical particles in suspensions
  confined between two planar walls}.
\newblock {\em J. Fluid Mech.}, in review:xxxx, 2004.

\bibitem{Liron-Mochon:1976}
N.~Liron and S.~Mochon.
\newblock {Stokes flow for a stokeslet between two parallel flat plates}.
\newblock {\em J. Engineering Math.}, 10:287--303, 1976.

\bibitem{Weinberger:1972}
H.~F. Weinberger.
\newblock {Variational properties of steady fall in Stokes flow}.
\newblock {\em J. Fluid Mech.}, 52:321--44, 1972.

\bibitem{Blawzdziewicz-Wajnryb-Given-Hubbard:2005}
J.~B{\l}awzdziewicz, E.~Wajnryb, J.~A. Given, and J.~B. Hubbard.
\newblock {Sharp scalar and tensor bounds on the hydrodynamic friction and
  mobility of arbitrarily shaped bodies in Stokes flow}.
\newblock {\em Phys. Fluids}, xxx:xxx, 2005.

\bibitem{Pesche-Nagele:2000}
R.~Pesch\'{e} and G.~N{\"{a}}gele.
\newblock {Stokesian dynamics study of quasi-two-dimensional suspensions
  confined between two parallel walls}.
\newblock {\em Phys. Rev. E}, 62:5432--43, 2000.

\bibitem{Lamb:1945}
H.~Lamb.
\newblock {\em Hydrodynamics}.
\newblock Dover, New York, 1945.

\bibitem{Faxen:1923}
H.~Faxen.
\newblock {}.
\newblock {\em Arkiv. Mat. Astron. Fys.}, 17:No. 27, 1923.

\bibitem{Happel-Brenner:1986}
J.~Happel and H.~Brenner.
\newblock {\em Low {Reynolds} Number Hydrodynamics}.
\newblock Martinus Nijhoff, Dordrecht, 1986.

\bibitem{Frenkel-Smit:2002}
D.~Frenkel and B.~Smit.
\newblock {\em {Understanding Molecular Simulation. From Algorithms to
  Simulations}}.
\newblock Academic Press, New York, 2002.

\bibitem{Blawzdziewicz-Vlahovska-Loewenberg:2000}
J.~B{\l}awzdziewicz, P.~Vlahovska, and M.~Loewenberg.
\newblock Rheology of a dilute emulsion of surfactant-covered spherical drops.
\newblock {\em Physica A}, 276:50--80, 2000.

\end{thebibliography}

\end{document}